\documentclass[preprint]{aastex}

\usepackage{verbatim}

\def\approx{$\sim$}
\def\persqcm{$\rm cm^{-2}$}
\def\percc{$\rm cm^{-3}$}
\def\h2{$\rm H_2$}
\def\error{$\pm$}
\def\e#1{$\times 10^{#1}$}
\def\tenup#1{10$^{#1}$}
\def\asec{\arcsec}

\def\deg{\arcdeg}
\def\thr{$^h$}
\def\tmin{$^m$}
\def\tsec{$^s$}

\def\kms{km~s$^{-1}$}

\def\solmass{$\rm M_{\sun}$}

\newcommand{\htooco}{$\rm H_2/CO$}
\newcommand{\convunits}{$\rm cm^{-2}\,(K\,km\,s^{-1})^{-1}$}

\newcommand{\jbks}{$\rm Jy\,b^{-1}\,km\,s^{-1}$}
\newcommand{\jykms}{$\rm Jy\,km\,s^{-1}$}
\newcommand{\mjb}{$\rm mJy\,b^{-1}$}
\newcommand{\re}{$r_e$}

\slugcomment{To appear in the {\it Astronomical Journal,} August 2002}

\begin{document}

\title{Molecular Gas in Elliptical Galaxies: Distribution and Kinematics}
\author{L. M. Young}
\affil{Physics Department, New Mexico Institute of Mining and Technology,
Socorro, NM 87801}
\email{lyoung@physics.nmt.edu}

\begin{abstract}  
I present interferometric images (\approx 7\asec\ resolution) 
of CO emission in five
elliptical galaxies and nondetections in two others.
These data double the number of elliptical galaxies whose CO emission
has been fully mapped.
The sample galaxies have \tenup{8} to 5\e{9} \solmass\ of molecular 
gas distributed in mostly symmetric rotating disks with diameters of 2
to 12 kpc.
Four out of the five molecular disks show remarkable
alignment with the optical major axes of
their host galaxies.
The molecular masses are a few percent of the total dynamical masses
which are implied if the gas is on circular orbits.
If the molecular gas forms stars, it will make rotationally supported
stellar disks which will be very similar in character to the stellar
disks now known to be present in many ellipticals.
Comparison of stellar kinematics to gas kinematics in NGC~4476 implies
that the molecular gas did not come from internal stellar mass loss
because the specific angular momentum of the gas is about three times
larger than that of the stars.

\end{abstract}


\section{Introduction}

It is now well known that elliptical galaxies often do have interstellar
media with some cold neutral gas and dust.
\citet{huchtmeier95} have found that about two thirds of ellipticals in
the RSA catalog contain HI at levels $M(HI)/L_B \geq 10^{-3}$ in solar
units; \citet{wardle86} reach similar conclusions.
\citet{colbert01} find that dust is apparent
in optical images of about 75\% of all ellipticals regardless of their
environment (field vs.\ X-ray detected poor groups).
The molecular gas content of ellipticals is more difficult to quantify
because, with few exceptions, only the ones which are bright in the
far-IR (FIR) have been searched.
However, \citet{knapp96} quote CO detection rates of 20 to 80\%
for ellipticals which are brighter than 1 Jy at 100\micron.

Since it was believed for many years that elliptical galaxies have little
or no cold molecular gas, detailed studies of that molecular gas can
offer fundamental insight into the evolution of ellipticals.
For example, one would obviously like to know the origin of the molecular
gas.  Did it come from internal sources (stellar mass loss) or from an
external source (another galaxy)?  Has it been there for a Hubble time or
significantly less?
The distribution and kinematics of the molecular gas, and particularly
comparisons of the specific angular momentum of the gas and the stars,
can help clarify the origin of the molecular gas.
Molecular gas is also the raw material for star formation. 
Therefore, the properties of the molecular gas determine where and how
much star formation can happen; this determines the future morphology of
the galaxy.
Finally, molecular gas distribution and kinematics are valuable because
the dissipational nature of gas means that the shapes of the gas orbits are
much better known than are the stellar orbits (e.g.
\citet{dezeeuw89, cretton00}).
Gas kinematics can be used to infer the galaxy potential in a way which
is more robust than, or at least complementary to, 
what one can do with stellar kinematics.
This paper uses high resolution CO observations to investigate
these ideas about elliptical galaxy structure and evolution.

Several authors have used single-dish telescopes to search for CO emission
from ellipticals.  The largest of these works are 
\citet{lees91}, \citet{wiklind95}, and \citet{knapp96}.
A very small number of  elliptical galaxies have been mapped in CO with
millimeter interferometers or with multiple pointings on single dish
telescopes.
These include NGC~759 \citep{wiklind97}, 
NGC~1275 \citep{reuter93, braine95, inoue96},
NGC~7252 \citep{wang92},
NGC~1316 \citep{horellou01}, and 
NGC~5128 = Cen~A \citep{quillen92, rydbeck93, charmandaris00}.
In all of these galaxies except NGC~7252, molecular gas is found in a
rotating disk on the order of a kpc or a few kpc in radius and containing
about \tenup{9} \solmass\ of H$_2$.
In Cen~A, a nearby galaxy which permits detailed observations,
the large scale CO disk closely follows the prominent optical dust lane
and is strongly warped \citep{quillen92}.  About 10\% of the CO in
Cen~A
is associated with stellar and HI shells at galactocentric radii of 15 kpc
\citep{charmandaris00}.
The CO in NGC~7252, a merger remnant, has compact but irregular structure
and kinematics.

The present paper doubles the number of elliptical galaxies with CO maps;
I show images of CO emission in five elliptical
galaxies and nondetections in two others.
In addition, the present sample is valuable because it employs a
clearly-defined set of selection criteria (in contrast to the
semi-random collection of interesting galaxies mentioned above).

\section{Sample selection}

The observed galaxies were chosen from a survey of CO
emission in ellipticals that was made with the IRAM 30m telescope by
\citet{wiklind95} [WCH].
WCH, \citet{lees91}, \citet{gordon91}, \citet{sage89}, \citet{knapp96},
and several other sets of authors 
selected galaxies for single dish CO 
surveys based on a combination of IRAS 60 $\mu$m and 100 $\mu$m fluxes and galaxy
type.
The most common FIR flux criterion (used by WCH and all of the surveys mentioned
here except \citet{sage89}) is $S_{100 \mu m} > 1.0$ Jy, where the 100
\micron\ fluxes were taken from the compilation of \citet{knapp89}.
WCH attempted to pick out ``genuine" ellipticals by restricting their sample to
galaxies known to have an $r^{1/4}$ profile or, in their words, ``a consistent
classification as E in several catalogs.''
Those criteria defined a sample of 29 ellipticals, of which 16 were
detected in CO.

The present sample contains all but one of the galaxies that were detected by WCH with
$^{12}$CO 1-0 integrated intensities greater than 5.0 K~\kms\ 
(23 \jykms) 
and that lie within the declination range accessible to BIMA and OVRO.
NGC~759, which also meets these criteria, was excluded because a high resolution CO map 
of this galaxy has already been published \citep{wiklind97}.
To this list I also added NGC~4649, for which a CO detection is reported by 
\citet{sage89}.
The resulting sample is given in Table \ref{sampletable}.

\begin{deluxetable}{lccrccl}
\tablewidth{0pt}
\tablecaption{Sample Galaxies
\label{sampletable}}
\tablehead{
\colhead{Galaxy} & \colhead{RA} & \colhead{Dec} & \colhead{Velocity} &
\colhead{Distance}  & \colhead{L$_B$} & \colhead{Environment} \\
\colhead{}       & \multicolumn{2}{c}{J2000.0} & \colhead{km/s} &
\colhead{Mpc} & \colhead{\tenup{10} L$_\odot$} & \colhead{}
}
\startdata
UGC 1503 & 02 01 19.8 & +33 19 46 & 5086 (6) & 69 & 1.7 & field \\
NGC 807  & 02 04 55.7 & +28 59 15 & 4764 (12) & 64 & 3.2 & field \\
NGC 3656 & 11 23 38.4 & +53 50 31 & 2869 (13) & 45 & 1.6 & merger remnant\\
NGC 4476 & 12 29 59.2 & +12 20 55 & 1978 (12) & 18 & 0.35 & Virgo cluster \\
NGC 4649 & 12 43 39.6 & +11 33 09 & 1117 (6) & 18 & 6.7 & Virgo cluster \\
NGC 5666 & 14 13 09.1 & +10 30 37 & 2221 (6)  & 35 & 0.63 & field \\
NGC 7468 & 23 02 59.3 & +16 36 19 & 2081 (6)  & 28 & 0.39 & group \\
\enddata
\tablecomments{Velocities are taken from the NASA Extragalactic Database
(NED); they refer to HI where available or to stellar velocities.  
The distance estimates are taken from WCH, who used 
$H_0 = 75$ \kms~Mpc$^{-1}$ and a Virgocentric infall
model which is described in their paper.  
The members of the Virgo cluster were
assumed to be at 18 Mpc.  Blue luminosities are taken from integrated
magnitudes in RC3; environmental descriptors are taken from WCH.
}
\end{deluxetable}

There is significant overlap between the sample selected here from the
survey of WCH and other single dish CO surveys.
NGC~5666 and NGC~4476 have the second and third highest CO 2-1 intensities 
in the sample of 24 galaxies studied by \citet{lees91}.
NGC~5666 has the highest CO 2-1 intensity in the sample of seven ellipticals
studied by \citet{gordon91}.

The galaxies observed by WCH are found almost evenly divided among the field,
groups, and clusters.  
Of the galaxies observed here, three are classified by WCH as being field
ellipticals, one is a member of a small group, two are in the
Virgo cluster, and one is most likely a merger remnant.
If molecular gas is associated with dust, then
this distribution of environments is consistent with the 
study of \citet{colbert01},
who found that optical signatures of dust are found in field ellipticals
at the same rate as in ellipticals in X-ray bright groups. 
Additional discussion of selection effects in the present sample can be
found in section \ref{caution}.

\section{Observations and data reduction}

\subsection{BIMA data}

Six galaxies [UGC~1503, NGC~807, NGC~3656, NGC~4476, NGC~4649, and
NGC~5666] were observed with the 10-element Berkeley-Illinois-Maryland
Association (BIMA) millimeter interferometer at Hat Creek, CA
\citep{welch96}.
The BIMA observations were carried out in the C configuration (projected 
baselines 3 to 34 k$\lambda$) between November 1998 and June 2001.
One additional track in the D configuration 
was obtained for NGC~5666 in March 1999, giving projected baselines
down to 2.3 k$\lambda$ for that galaxy. 

Each galaxy was observed with a single pointing centered on the optical
center of the galaxy;   
the primary beam FWHM is about 100\asec.
Each observation covered a velocity range of about 1000 \kms\ centered on
the velocity of the CO detected by WCH.
The optical velocities of the galaxies are uncertain by up to 100 \kms\
but are always well within the velocity range covered.
Table~\ref{sampletable} gives some basic data for the sample galaxies
and Table~\ref{obstable2} summarizes important parameters of the
observations and the final images.

\begin{deluxetable}{lccrrccc}
\rotate
\tablewidth{0pt}
\tablecaption{Observation and Image Parameters
\label{obstable2}}
\tablehead{
\colhead{Galaxy} & \colhead{Observation dates} & \colhead{Fluxcal} &
\colhead{Velocity Range} &
\colhead{Beam} & \colhead{Linear res.} & 
\colhead{Channel} & \colhead{noise} \\
\colhead{} & \colhead{} & \colhead{} & \colhead{\kms} & \colhead{\asec} & \colhead{kpc} &
\colhead{\kms} & \colhead{\mjb}
}
\startdata
UGC 1503 & 2000 Nov -- 2001 June & uranus & 4530--5580 & 
7.10$\times$6.26 & 2.4$\times$2.1 & 30 & 8.5 \\

NGC 807  & 2000 Dec -- 2001 May  & uranus & 4230--5280 & 
7.00$\times$6.34 & 2.2$\times$2.0 & 30 & 8.0 \\
 & & & & 
9.43$\times$8.46 & 2.9$\times$2.6 & 50 & 6.8 \\

NGC 3656 & 2000 Mar, Apr         & 3c273  & 2375--3335 & 
7.76$\times$6.16 & 1.7$\times$1.3 & 30 & 16 \\
 & & & & & & 20 & 19 \\

NGC 4476 & 2000 Mar, Apr         & 3c273/saturn  & 1480--2440 & 
8.29$\times$5.66 & 0.73$\times$0.49 & 30 & 11 \\
 & & & & & & 20 & 13 \\

NGC 4649 & 2000 May              & 3c273  & 450--1800 & 
8.22$\times$6.07 & 0.72$\times$0.53 & 30 & 24 \\

NGC 5666 & 1998 Nov -- 1999 Apr  & mars/3c279 & 1865--2555 & 
10.32$\times$7.63 & 1.7$\times$1.3 & 30 & 15 \\
 & & & & 8.27$\times$5.99 & 1.4$\times$1.0 & 30 & 17\\

NGC 7468 & 2001 Apr, May         & uranus/3c454.3 & 1628--2772 & 
6.54$\times$5.45 & 0.89$\times$0.74 & 20.8 & 13 \\
\enddata
\end{deluxetable}

Reduction of the BIMA data was carried out using standard tasks in the
MIRIAD package \citep{sault95}.
Electrical line length calibration was applied to most of the tracks,
with a few exceptions in cases where the measurement was too noisy to be
useful or where the line length was a very smooth function of time.
Data from an atmospheric phase monitor \citep{lay99} were used
to estimate the magnitude of amplitude decorrelation, as described by
\citet{regan01} and \citet{wong01}.
A small interferometer with a fixed 100 meter baseline measures the
rms path length difference in the signal from a commercial broadcast
satellite; those data are scaled to the observing frequency and are scaled by
projected baseline length raised to the 5/6 power 
to estimate the amount of amplitude decorrelation 
in the data \citep{akeson98}.
An rms path length difference of 300 $\micron$ on a 100 m baseline produces
an amplitude decorrelation of 0.82 for observations at 3mm wavelength \citep{akeson98},
but the longest baseline in the present data is 88 m.
Twelve tracks with rms path lengths less than 300 $\micron$
were not explicitly corrected for decorrelation because 
normal amplitude calibration can take out most of the decorrelation effect
\citep{wong01}.
Fourteen tracks with rms path lengths in the range 300--700
$\micron$ were corrected using the MIRIAD task {\it uvdecor}, which 
multiplies up the data amplitudes to correct for
decorrelation losses and decreases the weights of data with
large decorrelation losses.
Data with larger rms path lengths than 700$\micron$ were generally not used.
The worst track that was used had a median amplitude correction factor of
1.15.
Amplitude corrections were applied to
all observed sources and calibrators before the absolute flux calibration
was made.

Absolute flux calibration was based on observations of Uranus or Mars.
When suitable planets were not available, the secondary
calibrator 3C273, which is usually monitored several times per month, was
used (see Table \ref{obstable2}).
Phase drifts as a function of time were corrected by means of a
nearby calibrator observed every 30 to 40 minutes. 
Gain variations as a function of frequency were corrected by the online
passband calibration system;
inspection of the data for 3C273 indicate that residual
passband variations are on the order of 10\% or less in amplitude and 2\deg\ in
phase across the entire band.

The calibrated visibility data were weighted by the inverse square of the
system temperature and the inverse square of the amplitude 
decorrelation correction factor, then Fourier transformed.
No continuum emission was evident in the line-free channels of any galaxy
(Table \ref{conttable}).
The dirty images were lightly deconvolved with the Clark clean algorithm,
as appropriate for these compact, rather low signal-to-noise detections.
Integrated intensity and velocity field maps were produced by the masking 
method: 
the deconvolved image cube was smoothed along both spatial and velocity 
axes, and the smoothed cube was clipped at about 2.5$\sigma$ in absolute
value.
The clipped version of the cube was used as a mask to define a
three-dimensional volume in which the emission is integrated over velocity.
This masking method 
is described in greater detail by 
\citet{wong01} and by \citet{regan01}.

\begin{deluxetable}{cc}
\tablewidth{2.5in}
\tablecaption{Continuum Flux Density Limits
\label{conttable}}
\tablehead{
\colhead{Galaxy} & \colhead{3mm continuum} \\
\colhead{}       & \colhead{mJy}
}
\startdata
UGC 1503 & $<$5.0 \\
NGC 807  & $<$5.4 \\
NGC 3656 & $<$13 \\
NGC 4476 & $<$7.0 \\
NGC 4649 & $<$12 \\
NGC 5666 & $<$13 \\
NGC 7468 & $<$5.3 \\
\enddata
\tablecomments{Continuum images were made by averaging all of the
line-free channels in the final image cubes.  The values quoted
here are 3 times the rms noise in the continuum images, so they 
should be interpreted as flux density limits for point sources at the
centers of their host galaxies.
}
\end{deluxetable}

\subsection{OVRO data}

One galaxy, NGC~7468, was observed with the 6-element Owens Valley Radio
Observatory (OVRO) millimeter interferometer \citep{padin91}.
Those data were obtained in the C, L, and E configurations (projected
baselines 4 to 44 k$\lambda$) during April and May 2001.
A single pointing was made on the optical center of the galaxy; the primary
beam FWHM was about 65\asec.
The correlator was set up with four modules of 32 channels, each channel 4 MHz wide;
the modules were overlapped to cover a total bandwidth of 464 MHz.
The data were calibrated using the MMA package \citep{scoville93}.
Absolute flux calibration was based on observations of Uranus;
the passband and time-dependent phase calibration used the nearby source 3C454.3.
The calibrated data were mapped in AIPS using ``natural'' weighting.
Subsequent image analysis was identical to that for the BIMA data.

\section{Results}

\subsection{Nondetections}

\begin{figure}
\includegraphics[clip,scale=0.4,angle=-90]{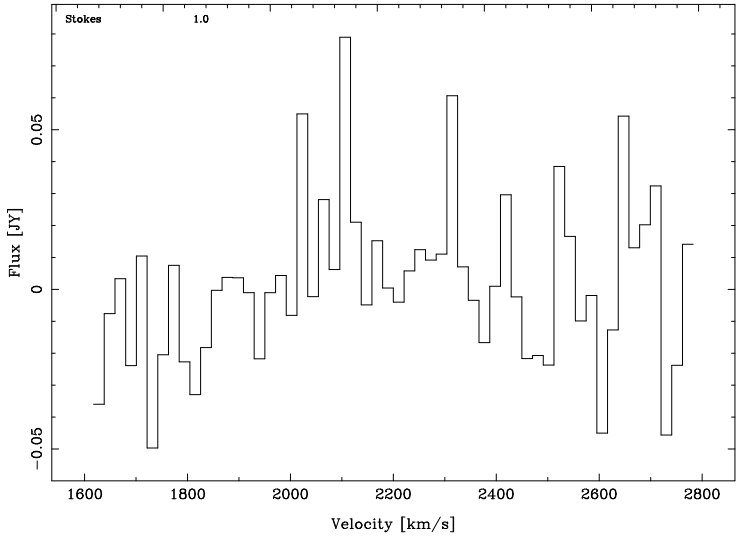}
\caption{CO spectrum of a square region, 22.5\asec\ on a side, centered
on the optical center of NGC~7468.
\label{7468spectrum}
}
\end{figure}

\begin{figure}
\includegraphics[clip,scale=0.4,angle=-90]{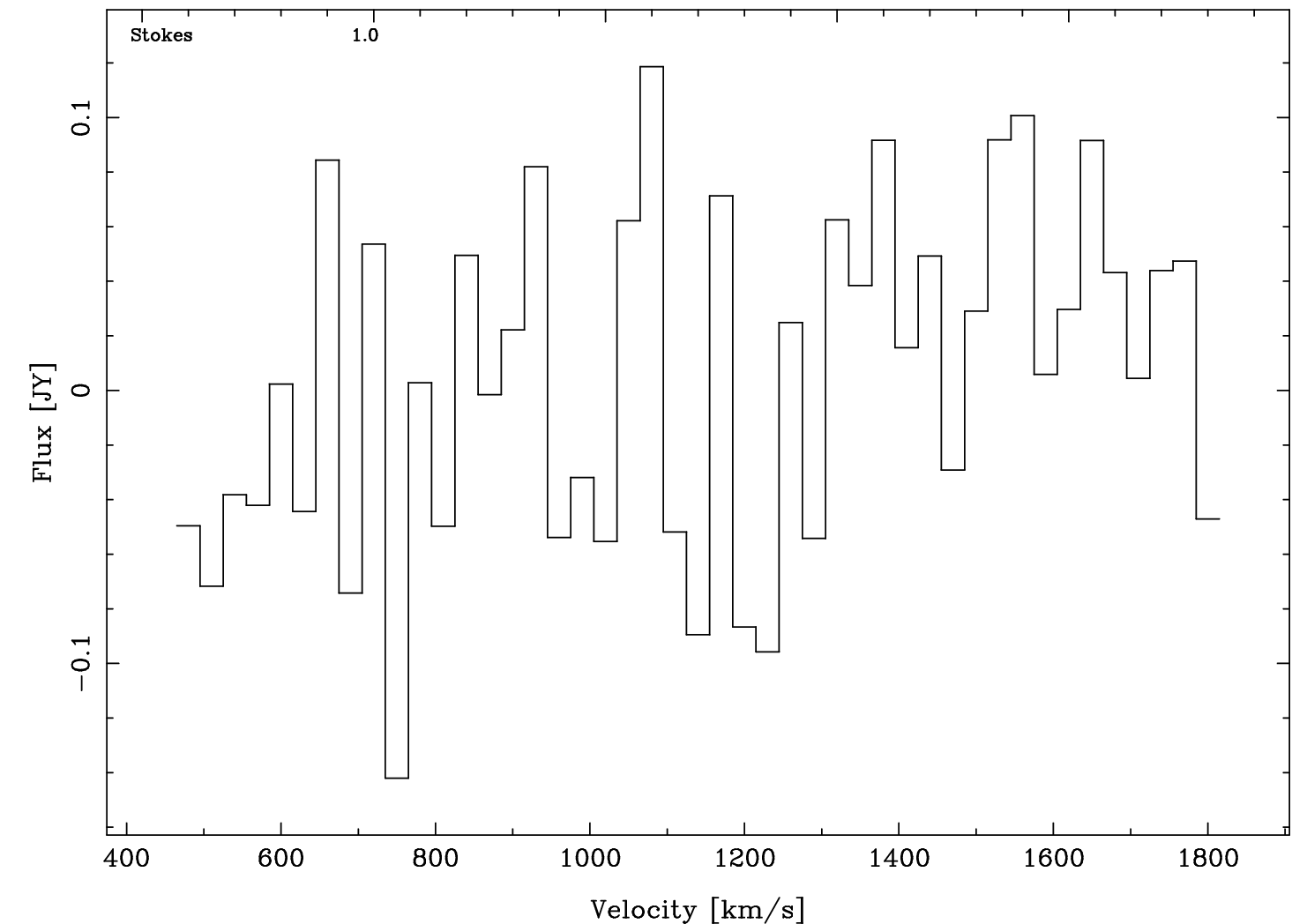}
\caption{CO spectrum of a square region, 22.5\asec\ on a side, centered
on the optical center of NGC~4649.
\label{4649spectrum}
}
\end{figure}

No CO emission was detected in the data cubes for NGC~4649 or NGC~7468.
Upper limits to the CO fluxes from these galaxies
were determined by first summing the data cube over a
square region 22.5\asec\ on a side, centered on the optical center of 
the galaxy, to
produce a spectrum.  
The 22.5\asec\ region was chosen to be similar in size and area to the
beam of the IRAM 30m telescope.
A spectrum was also produced for a 55\asec\ square region (similar to 
the area of the NRAO 12m telescope) for NGC~4649.
Nothing but noise is apparent in the spectra (Figures \ref{7468spectrum}
and \ref{4649spectrum}).  
The spectra were then summed over the velocity ranges described below.
The uncertainty in the sum is calculated from the rms in the spectrum and the number of
channels summed, as described by \citet{young00} and \citet{lees91}; 3$\sigma$ limits are given
in Table \ref{resultstable}.
This estimate assumes the channels are uncorrelated, which is an excellent assumption for 
these data since no baseline or continuum emission needs to be subtracted.

\begin{deluxetable}{lcccccrr}
\tablewidth{0pt}
\tablecaption{H$_2$ Mass and Morphology
\label{resultstable}}
\tablehead{
\colhead{Galaxy} & \colhead{CO flux} & 
\colhead{M(H$_2$)} &
\multicolumn{2}{c}{CO diameter} & 
\multicolumn{2}{c}{CO shape} & \colhead{Kin Maj Ax} \\
\colhead{}  & \colhead{\jykms}      & 
\colhead{\tenup{8} \solmass} &
\colhead{\asec} & \colhead{kpc} &
\colhead{$\epsilon$} & \colhead{PA ($\deg$)} & \colhead{PA ($\deg$)} 
}
\startdata
UGC 1503 & 32 (6)  & 18 & 30 & 11 (1) &  0.37 (0.05) & $-$111 (1) &
         $-$123 (2)  \\
NGC 807  & 29 (6)   & 14 & 40 & 12 (1) & 0.58 (0.09) & 143 (1) & 
        149 (3)  \\
NGC 3656 & 200 (20) & 47 & 34 & 7.4  (0.7) & 0.73 (0.02) & 174.8 (0.1) & 
           191 (3) \\
NGC 4476 & 30 (3) & 1.1 & 27 & 2.4 (0.2) & 0.57 (0.04) & $-$151 (1) &
            $-$152 (1) \\
NGC 4649 & $<$18 & $<$0.68 & \nodata & \nodata & \nodata & \nodata &
             \nodata \\
NGC 5666 & 39 (4)  & 5.7 & 28 & 4.7 (0.5) & 0.11 (0.06) & 166 (5) &
            165 (2) \\
NGC 7468 & $<$6.2 & $<$0.57 & \nodata & \nodata & \nodata & \nodata &
         \nodata \\
\enddata
\tablecomments{Upper limits are 3$\sigma$ for a 22.5\asec\ square region
and 300 \kms\ velocity range.  See text for further information.
Ellipticities and position angles (both morphological and kinematic) for 
the CO distributions are the median
values from fits to three or four images made at different resolutions or
with different clip levels.  The values in parentheses for these last
three columns are estimates of the uncertainty based on the spread among
the different fits, because those spreads are always larger than the
formal uncertainties of the fits.  Position angles are measured North
through East to the receding major axis.}
\end{deluxetable}

WCH reported a detection, which they characterize as tentative, of 6.7
K~\kms\ $\sim$ 31 \jykms\ in the $^{12}$CO 1-0 line from NGC~7468.  
Their line is centered at 2300 \kms\ (220 \kms\ distant from the optical
velocity) and 665 \kms\ wide.
The present OVRO observations give a sum of 4.7\error 2.9 \jykms\ 
over the 22.5\asec\ square region and over the same velocity range as the
CO line of WCH. 
The OVRO data also give a sum of 3.6\error 2.0 \jykms\ over the same spatial
region but over a 300 \kms\ velocity range centered on the HI velocity.
(The HI line in NGC~7468 is about 200 \kms\ wide at 20\% of peak
intensity; see \citet{vandriel00}.)
Emission as strong as that reported by WCH should have been easily detected.
I also consider it unlikely that CO in NGC~7468 is invisible to the
interferometer by virtue of being smoothly distributed; 
the 23\asec\ beam of the 30m telescope is only three times larger than the
7\asec\ beam of the OVRO data, so the relevant spatial scales are well sampled
in the interferometer data.

Thus, the OVRO data for NGC~7468 confirm the nondetection of the 2-1 line of CO by
\citet{lees91}.  Those authors quote an upper limit H$_2$ mass of
8\e{7} \solmass\ (after scaling to the conversion factor and distance
assumed here), which is consistent with the present 6\e{7} \solmass\ limit.
Details of the H$_2$ mass estimates are described in Section
\ref{fluxes}.

\citet{sage89} report the detection of 18.7\error 2.3 \jykms\ of
emission in $^{12}$CO 1-0 from NGC~4649.  Those observations were made
with the NRAO 12m telescope, which has a beam of 55\asec; the reported
line is 225 \kms\ wide.  
The detection is not considered tentative but again there is a rather large
offset (200 \kms) between the CO velocity and the optical velocity.
A sum over a 55\asec\ box and over the same velocity range noted by
\citet{sage89} gives an integrated flux of 33\error 18 \jykms\ in the present
images.
A sum of the BIMA data over a 22.5\asec\ box and over a 300 \kms\ range centered on the
optical velocity of the galaxy gives an integrated flux of $-$5.6\error
6.0 \jykms.
If the CO detected by \citet{sage89} was smoothly distributed over the
225 \kms\ velocity range and over the 55\asec\ beam of the 12m telescope,
it would be too faint to be detectable in the current interferometer
images.
Furthermore, it would be invisible to the interferometer, which does not
sample those 55\asec\ spatial scales.
If, however, the CO detected by \citet{sage89} was concentrated within
one $8''\times6''$ beam for each individual channel, it would have been
detectable at the 4$\sigma$ to 8$\sigma$ level.
Additional mosaic observations of NGC~4649 in BIMA's D configuration have better
sensitivity to large scale structures; those observations will be
reported in a future paper, and they also fail to detect CO in
NGC~4649.

\subsection{CO fluxes and H$_2$ masses}\label{fluxes}

For the galaxies with CO detections, total fluxes were measured from the
integrated intensity images shown in figures \ref{1503stars+co},
\ref{807stars+co}, \ref{3656stars+co}, \ref{4476stars+co}, and
\ref{5666stars+co}.
The uncertainties in the CO fluxes are probably 10\% for the stronger detections
(NGC~3656, NGC~4476, and NGC~5666),
dominated mostly by the absolute calibration.  
For NGC~807 and UGC~1503
the uncertainty is probably a bit larger, perhaps 15\% to 20\%, due to 
uncertainties in the absolute calibration and in choosing the spatial
region to be summed.

The CO fluxes measured in the present images of 
UGC~1503, NGC~807, and NGC~4476 are consistent within 20\% of the 
$^{12}$CO 1-0 fluxes measured by WCH using the IRAM 30m telescope.  
WCH detected a larger CO flux from NGC~5666, 60 \jykms, than I find in
the BIMA image (40 \jykms).  
The difference is nominally larger than the combined uncertainties, which
are about 10\% for the BIMA image and probably 10--15\% for the data
of WCH (C.\ Henkel, priv.\ comm.). 
However, there is no compelling evidence that the interferometer has
missed a significant component of the molecular gas in this galaxy.
Conversely, the images shown here reveal that the molecular gas
distributions for UGC~1503, NGC~4476, and NGC~5666 are not very much
larger than the 23\asec\ (FWHM) beam of the 30m telescope, so there is no
compelling evidence that the single dish spectra missed significant
components of the molecular gas in these galaxies.
The exception to this latter statement is NGC~3656, which has a factor
of two larger flux in the BIMA images than in the spectrum of WCH;
most likely this is because the gas distribution is larger than the
30m beam.
The CO line widths agree well in all cases.

\h2\ masses (Table \ref{resultstable})
are calculated using the distances in Table \ref{sampletable} and a
``standard" \htooco\ conversion factor of 3.0\e{20} \convunits\
as in WCH.  With this conversion
factor, \h2\ masses  are related to CO fluxes $S_{CO}$ by
$ M(H_2) = (1.18\times 10^4\: M_\odot)\: D^2\: S_{CO}$
where $D$ is the distance in Mpc and $S_{CO}$ is the CO flux in \jykms.
No correction has been made for the presence of helium.

\subsection{CO Morphology}
\label{COmorph}

\begin{figure}
\includegraphics[scale=0.8]{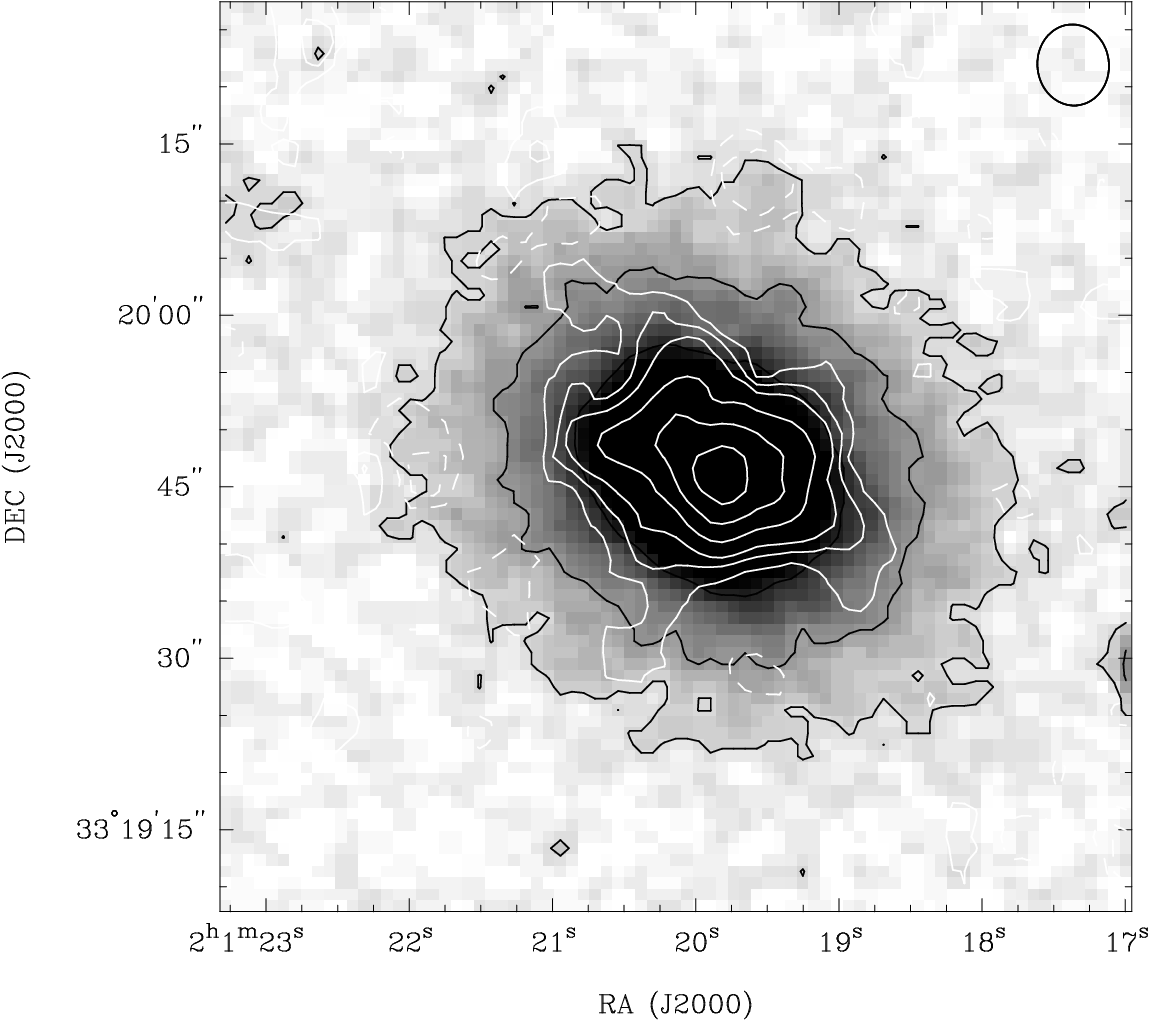}
\caption{Molecular gas in UGC~1503.  The greyscale and black
contours are an optical image from the red portion of the second
generation Digitized Sky Survey.  The heavy white contours show the CO
integrated intensity in units of $-20$, $-10$, 10, 20, 30, 50, 70, and 90
percent of the peak, which is 6.3 \jbks\ = 3.9\e{21} \persqcm\ using the
\htooco\ conversion factor described in Section \ref{fluxes}.  The small
ellipse at the top indicates the size of the beam.
\label{1503stars+co}
}
\end{figure}

\begin{figure}
\includegraphics[scale=0.8]{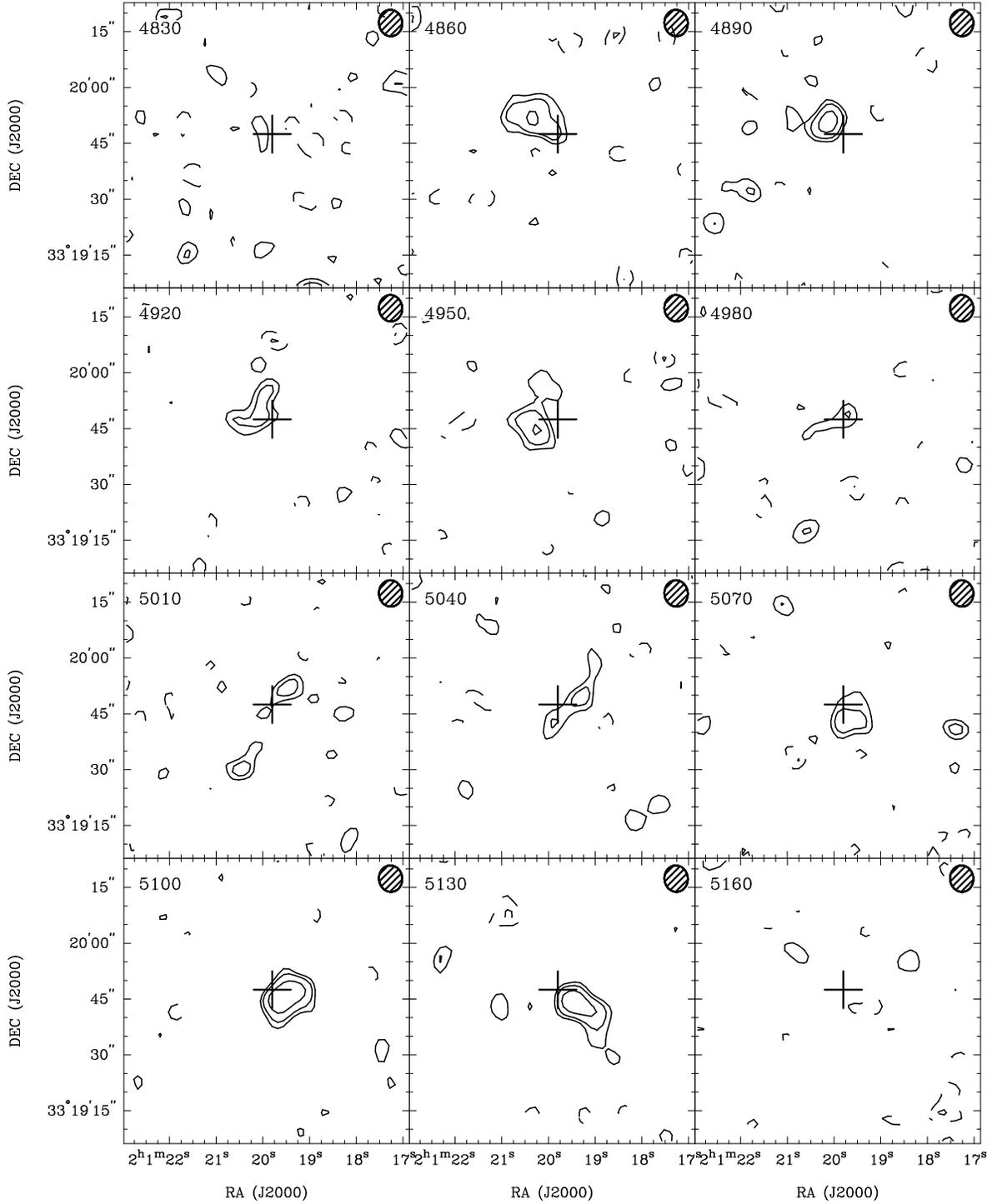}
\caption{Individual channel maps showing CO emission from UGC~1503.
Contour levels are $-$3, $-$2, 2, 3, 5, 8.3, 13.9, and 23.1 times 
8.0 \mjb\ \approx\ 1$\sigma$.
The velocity of each channel (in
\kms) is indicated in the upper left corner and the beam size in the
upper right corner.
The cross marks the kinematic center of the gas, which coincides with the
morphological center of the gas and the optical center to within 1\asec\ to
2\asec.
\label{1503channels}
}
\end{figure}

\begin{figure}
\includegraphics[scale=0.8]{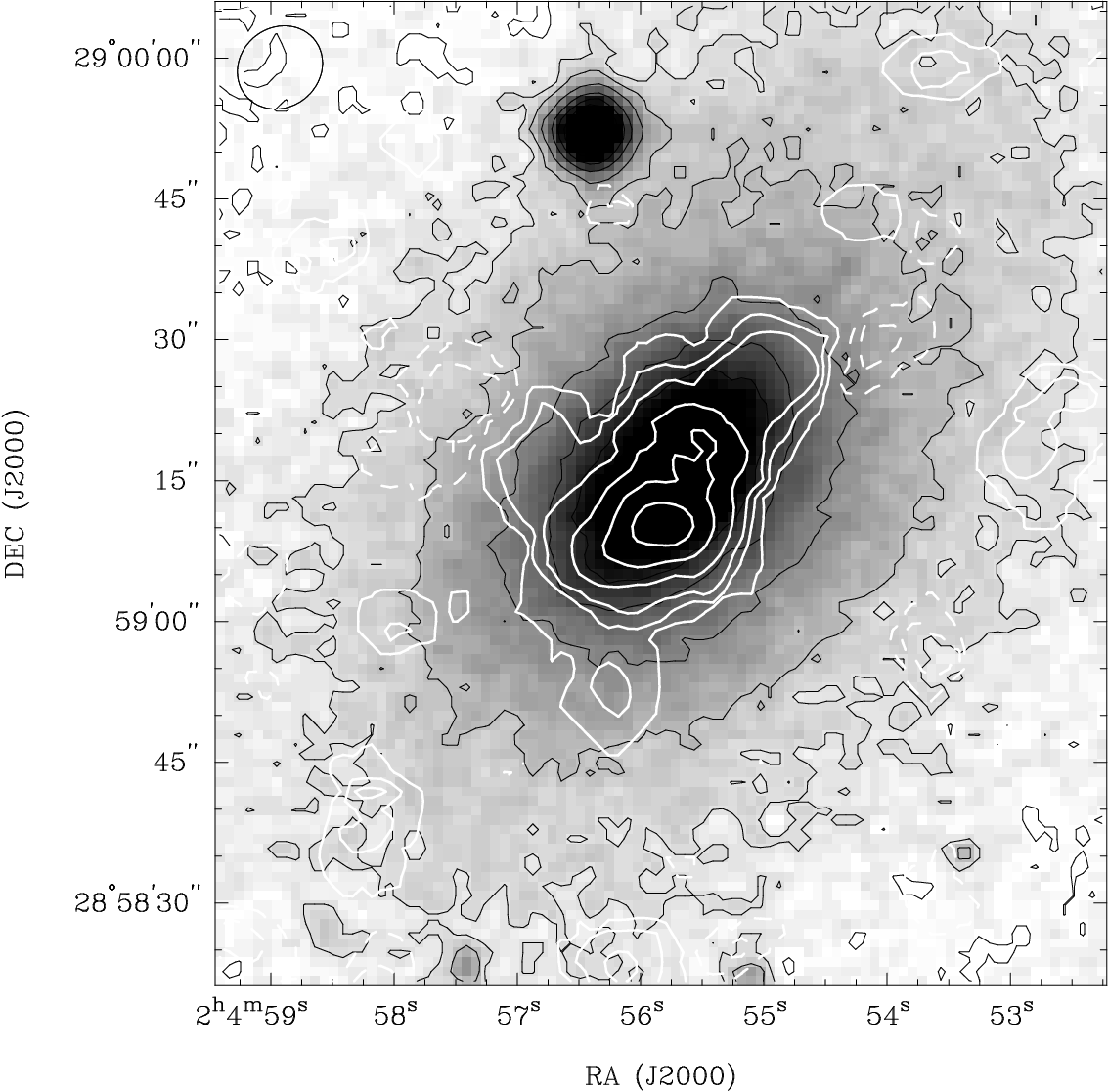}
\caption{Molecular gas in NGC~807.  Heavy white contours show the CO
integrated intensity in units of 
$-$20, $-$10, 10, 20, 30, 50, 70, and 90 
percent of the peak (7.6~\jbks\ = 2.6\e{21} \persqcm).
Other features as in Figure \ref{1503stars+co}.
\label{807stars+co}
}
\end{figure}

\begin{figure}
\includegraphics[scale=0.8]{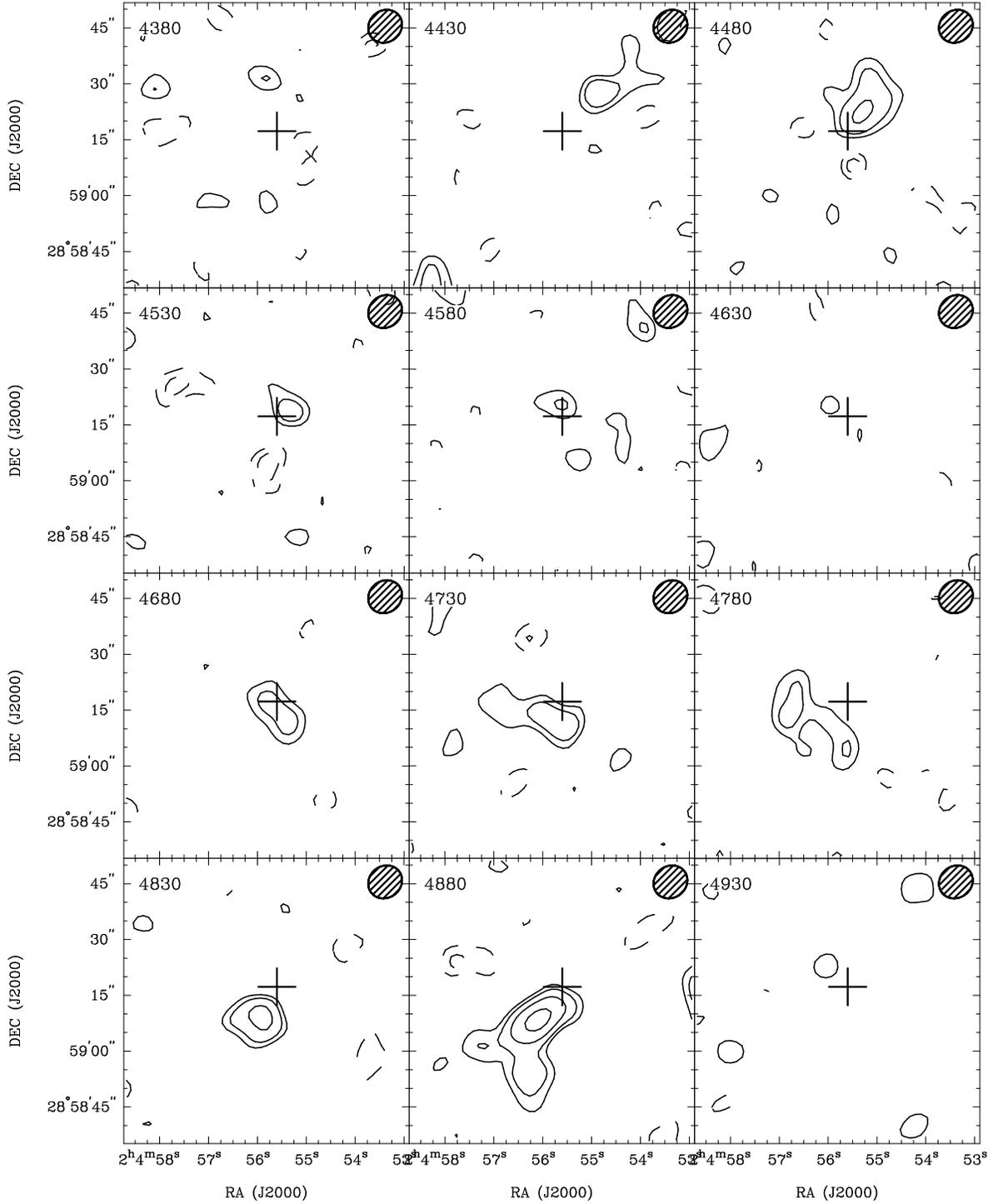}
\caption{Individual channel maps showing CO emission from NGC~807.
The contour intervals are the same as for Figure \ref{1503channels}
but the multiplicative unit is 1$\sigma$ = 6.8 \mjb.
The cross marks the kinematic center of the molecular gas;
that location is coincident with the optical center of the galaxy,
given the uncertainties (about 2\asec) in each
position.
\label{807channels}
}
\end{figure}

\begin{figure}
\includegraphics[scale=0.8]{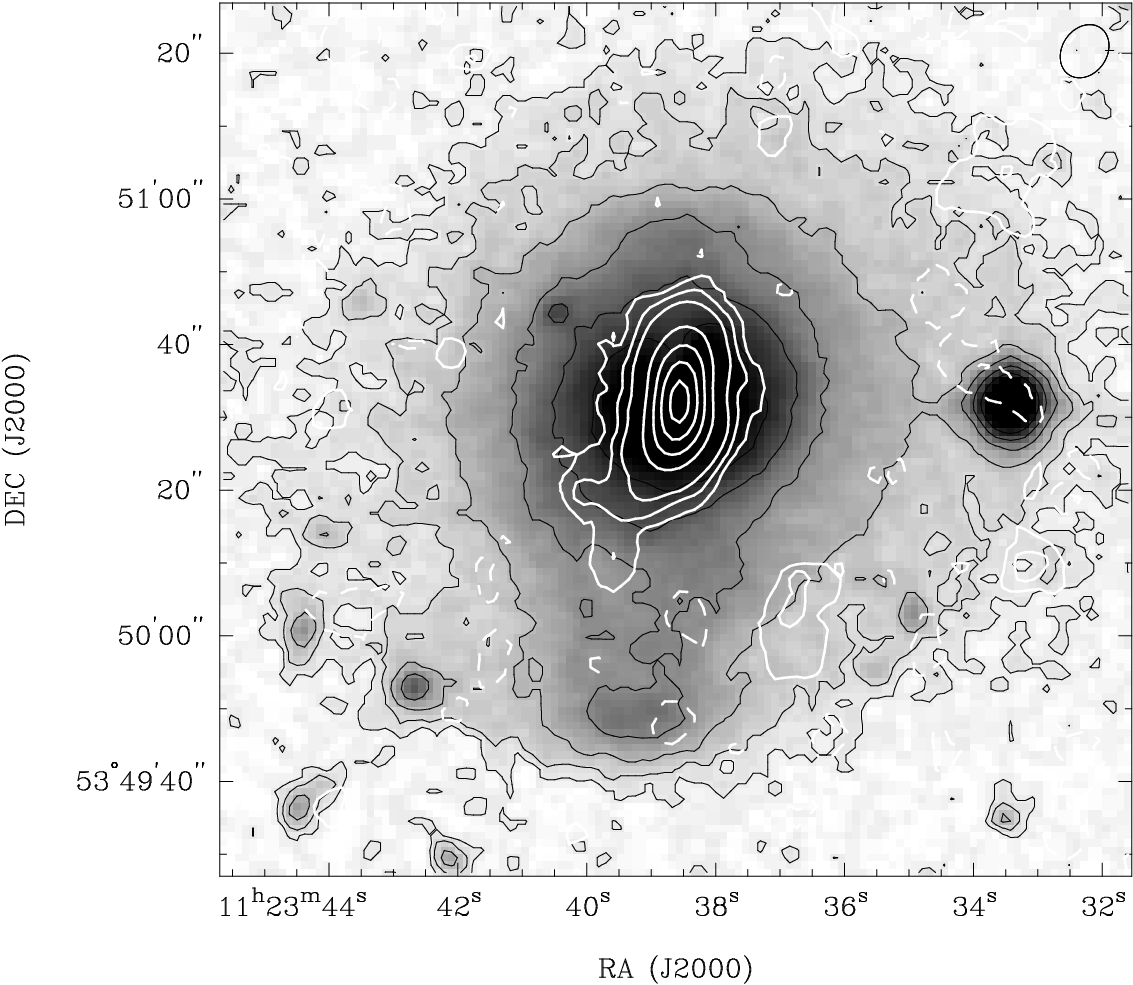}
\caption{Molecular gas in NGC~3656.  Heavy white contours show the CO
integrated intensity in units of $-5$, $-2$, 2, 5, 10, 20, 30, 50, 70, and 90
percent of the peak (81.1 \jbks\ = 4.7\e{22} \persqcm).  Other features as
in Figure \ref{1503stars+co}.
\label{3656stars+co}
}
\end{figure}

\begin{figure}
\includegraphics[scale=0.8]{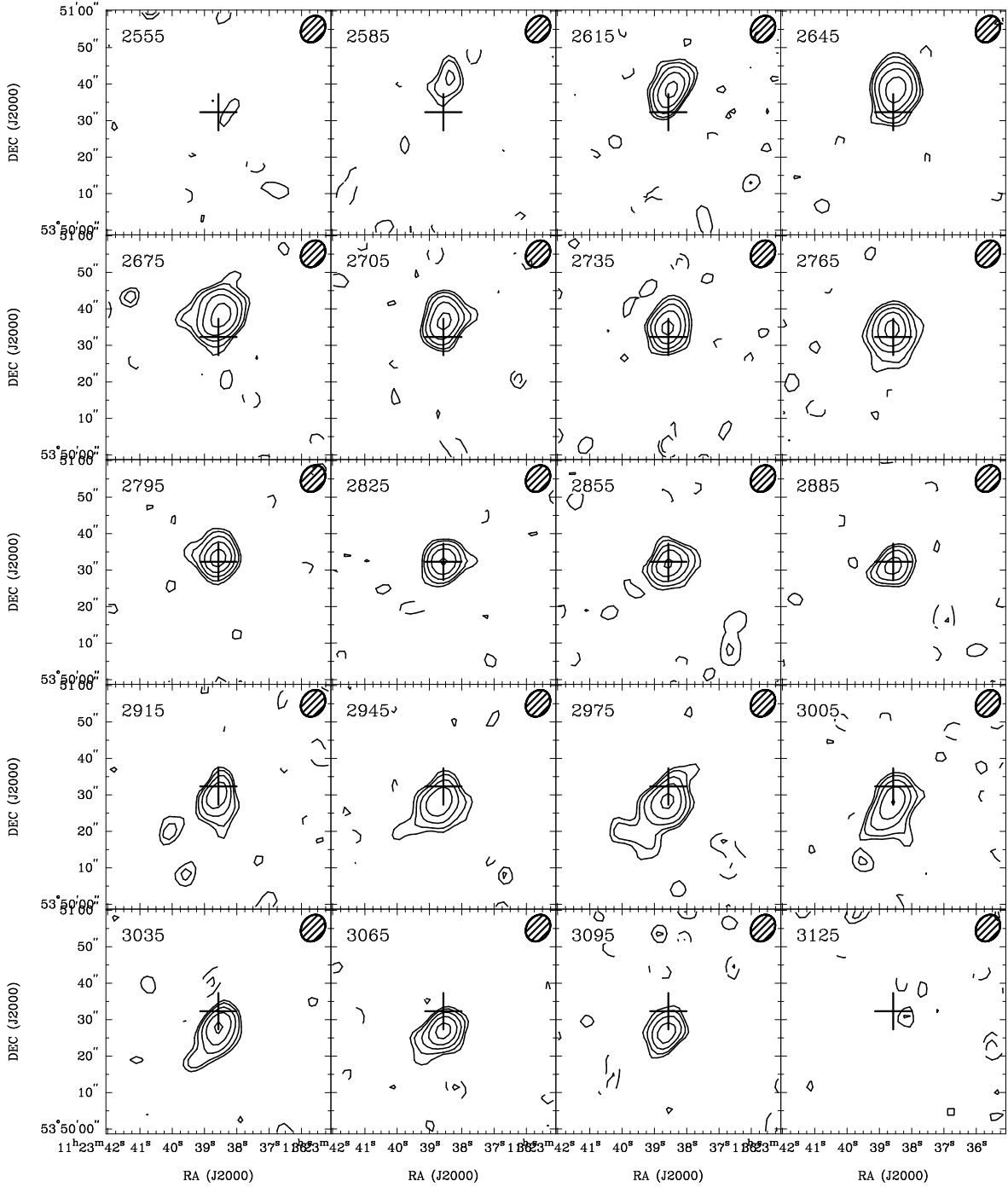}
\caption{Individual channel maps showing CO emission from NGC~3656.
As for Figure \ref{1503channels}, but the contour levels are multiplied by 
1$\sigma$ = 16 \mjb.
The cross marks the morphological center of the molecular gas, which is
coincident with the optical center to 1\asec.
\label{3656channels}
}
\end{figure}

\begin{figure}
\includegraphics[scale=0.8]{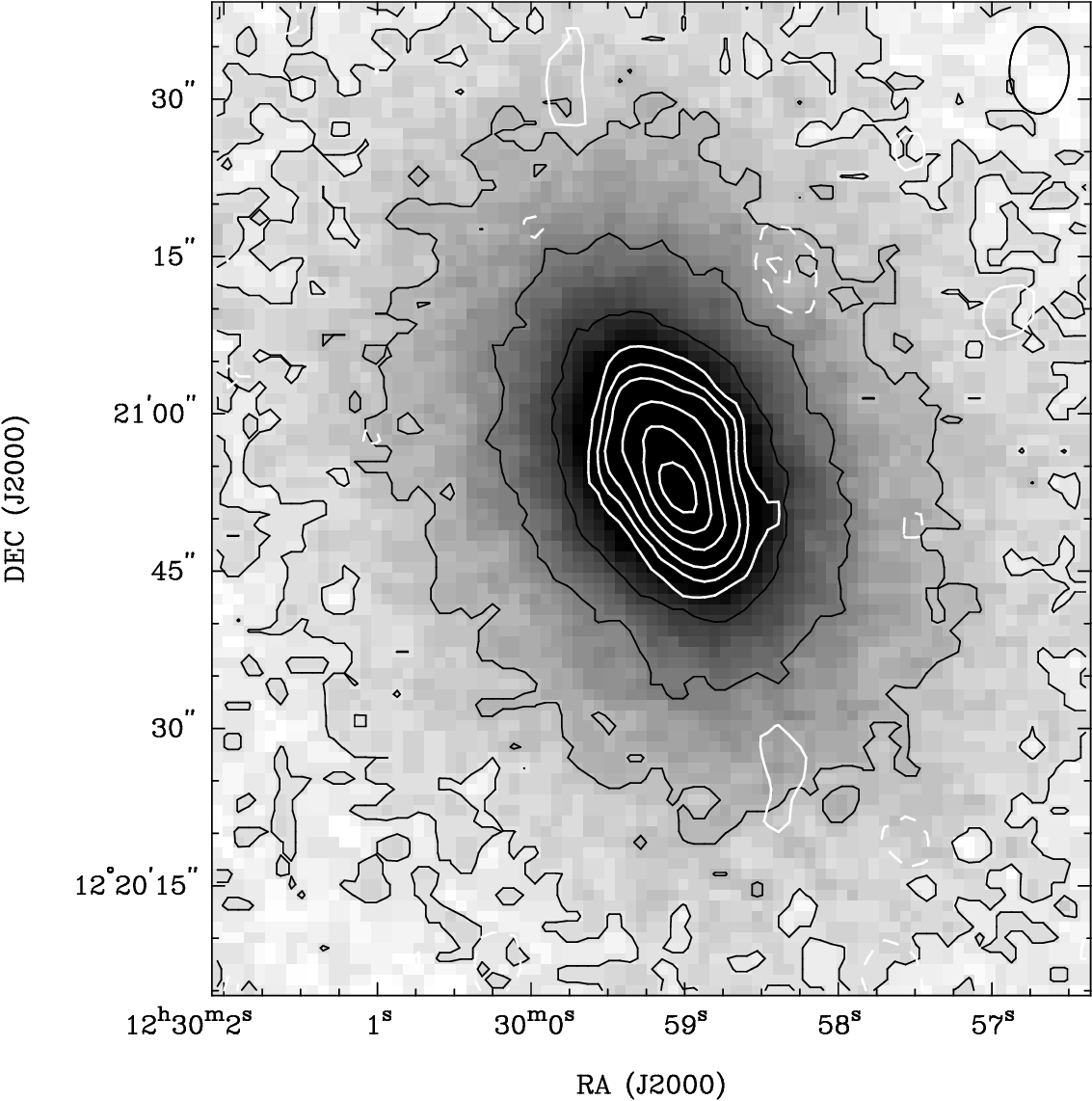}
\caption{Molecular gas in NGC~4476.  Heavy white contours show the CO
integrated intensity in units of $-20$, $-10$, 10, 20, 30, 50, 70, and 90
percent of the peak (12.4 \jbks\ = 7.3\e{21} \persqcm).  Other features as
in Figure \ref{1503stars+co}.
\label{4476stars+co}
}
\end{figure}

\begin{figure}
\includegraphics[scale=0.8]{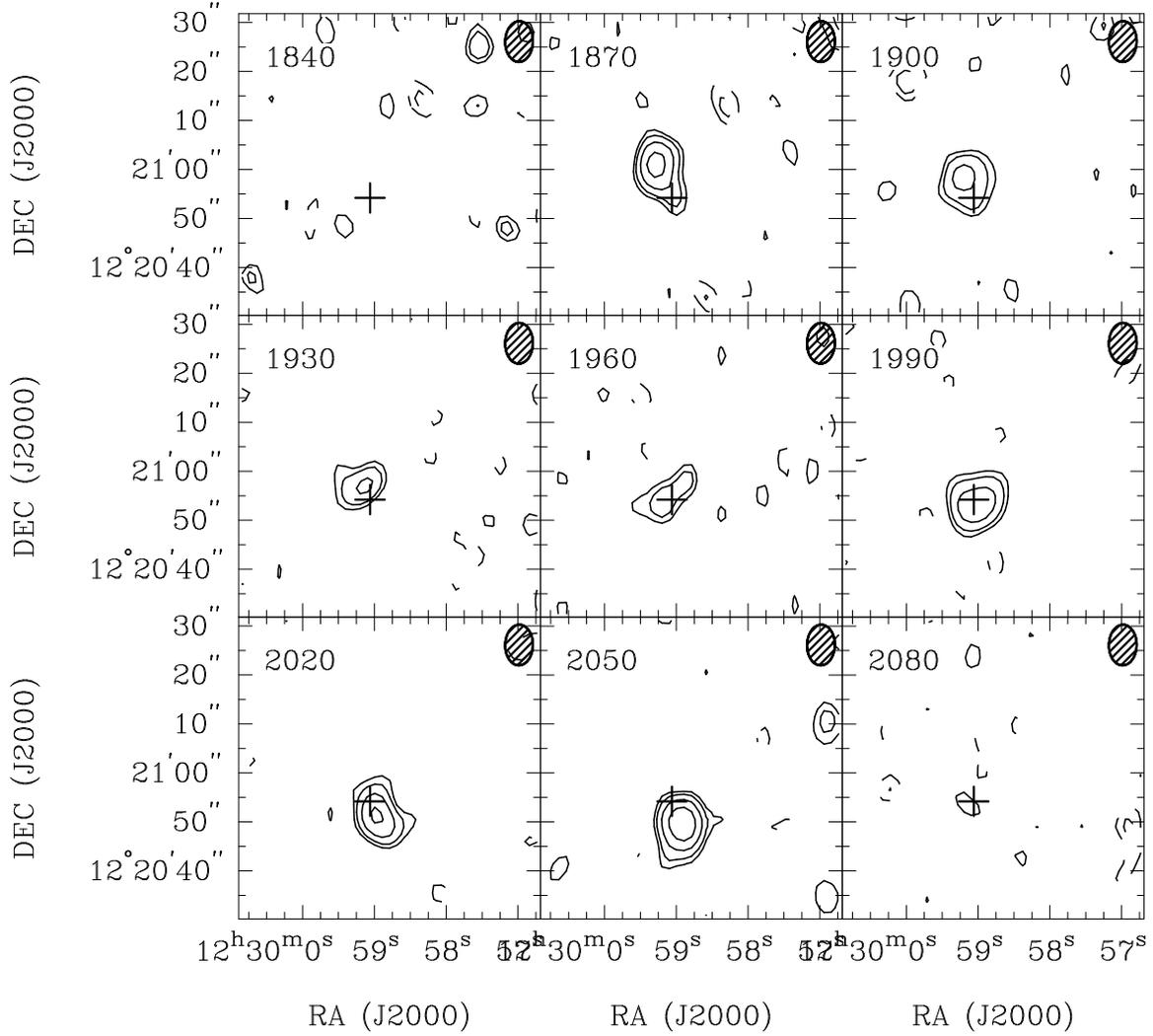}
\caption{Individual channel maps showing CO emission from NGC~4476.
As for Figure \ref{1503channels}, but the contour levels are multiplied by 
1$\sigma$ = 11.5 \mjb.
The cross marks the morphological center of the gas, which is coincident
with the optical center of the galaxy to about 2\asec.
\label{4476channels}
}
\end{figure}

\begin{figure}
\includegraphics[scale=0.8]{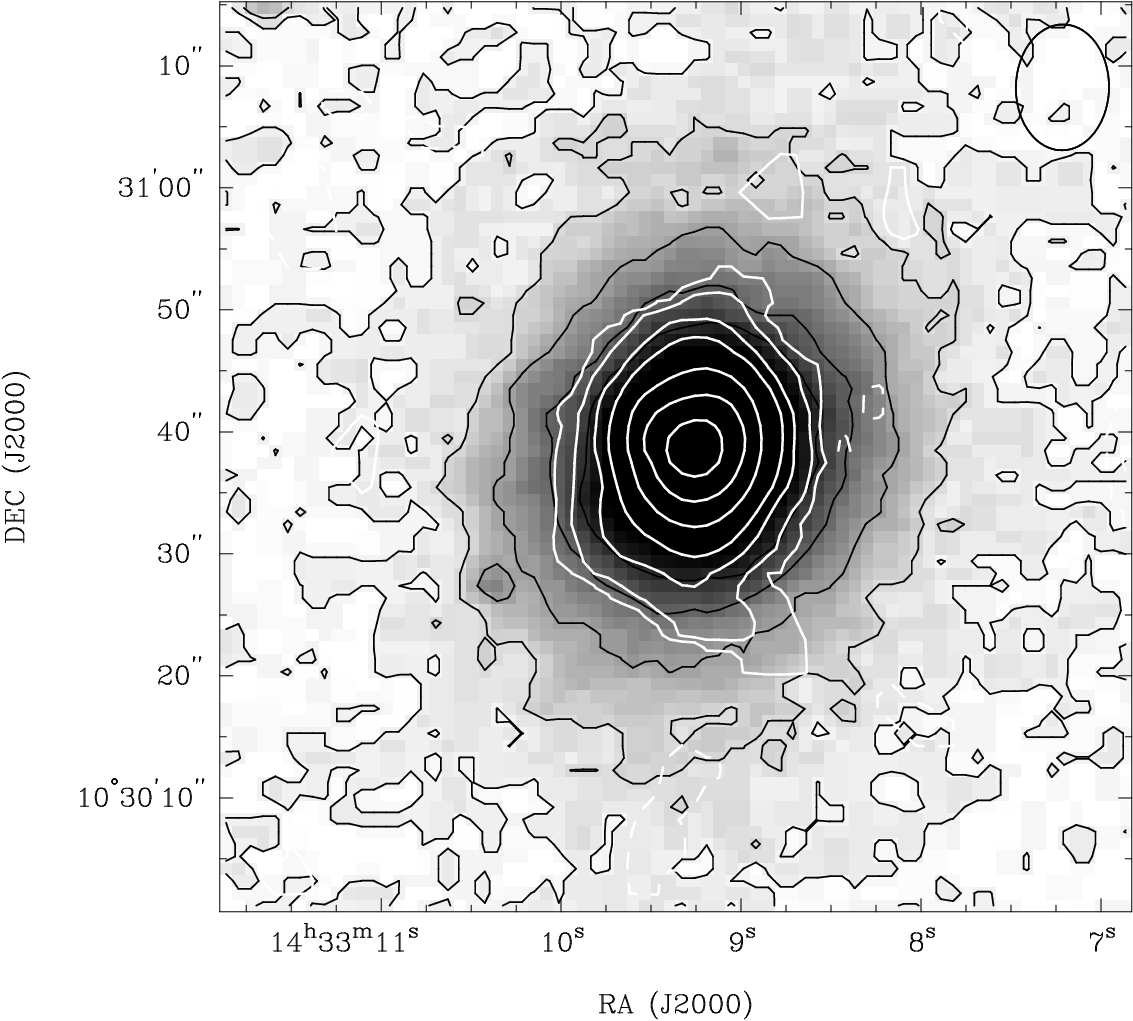}
\caption{Molecular gas in NGC~5666.  Heavy white contours show the CO
integrated intensity in units of $-10$, $-5$, 5, 10, 20, 30, 50, 70, and 90
percent of the peak (21.3 \jbks\ = 7.5\e{21} \persqcm).  Other features as
in Figure \ref{1503stars+co}.
\label{5666stars+co}
}
\end{figure}

\begin{figure}
\includegraphics[scale=0.8]{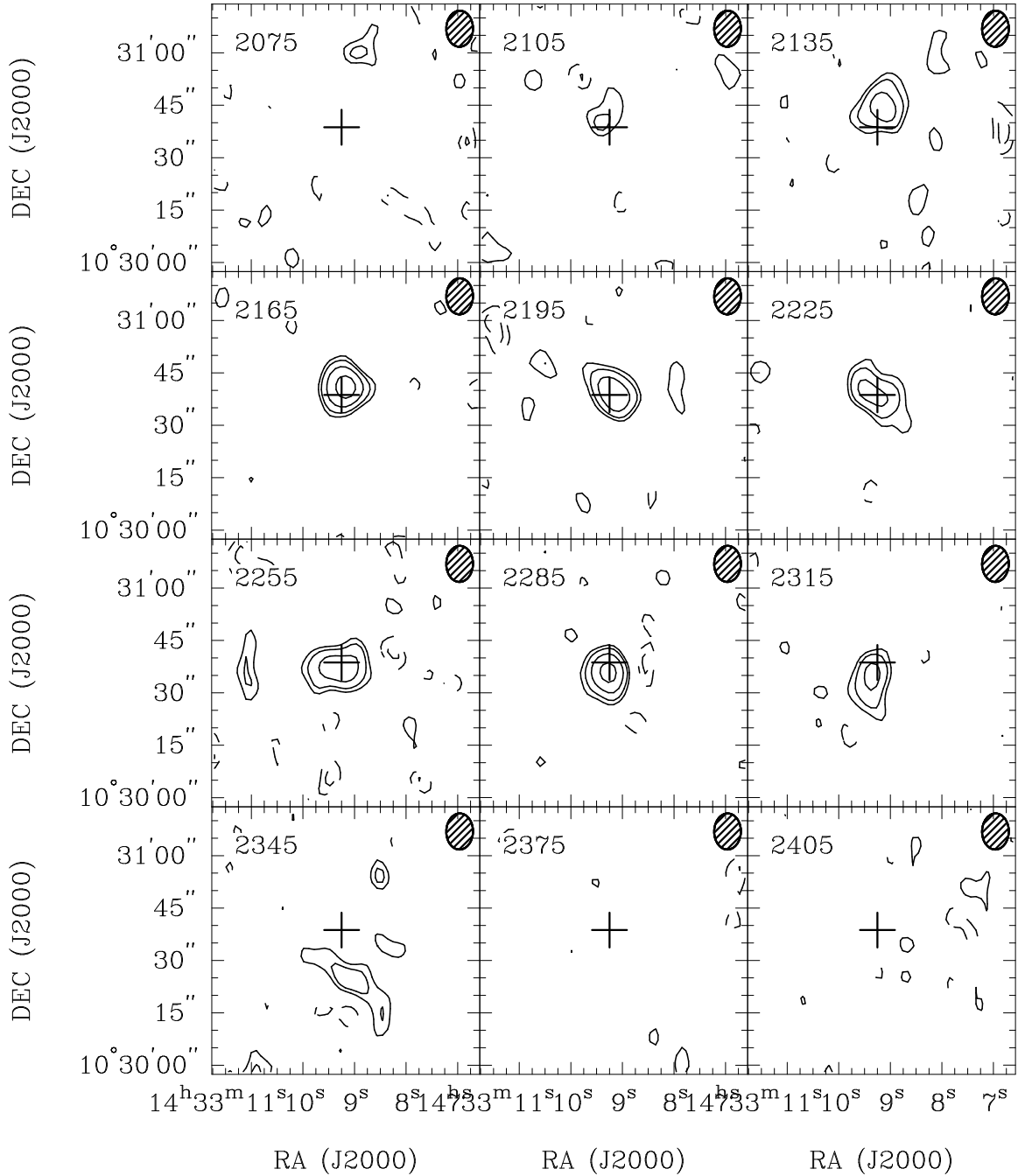}
\caption{Individual channel maps showing CO emission from NGC~5666.
As for Figure \ref{1503channels}, but the contour levels are multiplied by
1$\sigma$ = 15 \mjb.
The cross marks the morphological center of the molecular gas, which
coincides with the optical center given in NED to better than 1\asec.
\label{5666channels}
}
\end{figure}

Figures \ref{1503stars+co}, \ref{1503channels}, \ref{807stars+co},
\ref{807channels}, \ref{3656stars+co}, \ref{3656channels},
\ref{4476stars+co}, \ref{4476channels}, \ref{5666stars+co}, and
\ref{5666channels} show integrated CO intensity maps and 
channel maps for the five galaxies with detected CO emission.
The molecular gas in these five elliptical galaxies is found in very
regular, symmetric rotating disks with diameters of a few up to 12 kpc.
The disks appear flat at the current resolution and sensitivity;
the only feature which can be reliably identified outside of the
flat disks is in NGC~3656 (Figure \ref{3656stars+co}).
In this galaxy, the majority of the molecular gas is found in a disk
oriented north-south, following the optical dust lane \citep{balcells01}.
Roughly 6\% of the galaxy's total CO flux comes from a feature at
11\thr\ 23\tmin\ 40\tsec, 53\deg\ 50\arcmin\ 20\asec\ (J2000),  about
10\asec\ west of the southern end of the main CO disk.
The feature
is also visible in the individual channel images (Figure
\ref{3656channels}) near 2975 \kms.
The features above and below the disk of NGC~807 (Figure
\ref{807stars+co}) may simply be noise.

With the current rather low resolution it is difficult to be sure whether
the elongated molecular gas distributions come from disks or bars.
I assume in the majority of this paper that they are disks because 
of the characteristic ``butterfly" pattern which is apparent in 
the channel maps for UGC~1503, NGC~807, and NGC~4476 (Figures
\ref{1503channels}, \ref{807channels}, and \ref{4476channels}).
In this pattern, the channels near to the
systemic velocity show gas distributions which are elongated in the
direction of the kinematic minor axis but the edge channels show much
more compact gas distributions.
The pattern is characteristic of roughly circular gas disks inclined to
the line of sight.
NGC~807 and UGC~1503 also show rotation curves which rise and then
flatten in the manner common to spiral galaxy gas disks (Section
\ref{COkin}).
The butterfly pattern is not obvious in the channel maps for
NGC~3656 (Figure \ref{3656channels}); I will continue to assume that 
the gas is in a disk rather than a bar, but higher resolution observations
would be beneficial.

Two of the galaxies show evidence for asymmetries in their gas
distributions (the 
CO emission is stronger on one side than the other).
This effect is dramatic for NGC~807 (Figure \ref{807stars+co}), where approximately 30\% of the
CO emission comes from the northwest half of the galaxy and 70\% from the
southeast half.
The asymmetry is also apparent in the channel maps of Figure
\ref{807channels}, where the peak intensity in the channel at
4880 \kms\ is 10$\sigma$ but the peak intensity in the matching channels on
the other side of the systemic velocity (4430 and 4480 \kms) is only
5$\sigma$.  Such a large intensity difference is unlikely to be due
to noise.
WCH's spectrum of NGC~807 does not show the high velocity side to be
stronger than the low velocity side, but this is probably because the
peak intensity in the 4880 \kms\ channel occurs 10\asec\ away from
the galaxy center  (see also Figure \ref{807pv}).
This means that the strongest emission in the galaxy 
comes from regions at the half power
point of the IRAM 30m telescope beam, and the 30m telescope has much less
sensitivity to this feature than the interferometer.
A lopsided CO distribution is also evident in NGC~4476 (Figure
\ref{4476stars+co}), though to
a lesser extent. 

None of the galaxies show definitive evidence for CO at large radii as in
Cen~A \citep{charmandaris00}.  In that galaxy, at least
10\% of the CO emission of the galaxy is not associated with the 
optical dust lane but is found in stellar shells at 15 kpc radius.
The rms noise levels in the images (Table \ref{obstable2})  are such that
an unresolved source appearing in one channel at the 5$\sigma$ level
would have a CO flux of 1 to 2 \jykms, which is less than 10\% 
of the CO fluxes of all of the galaxies detected here (Table
\ref{resultstable}).
Thus, if 10\% of the CO emission were in features like those of Cen~A
it would most likely have been detected.
The numbers are particularly compelling for NGC~3656 (the galaxy which,
optically, looks most disturbed).  
Molecular gas associated with the optical shell and containing 
as little as 1\% of the total CO flux of that galaxy would probably have
been detected.
Thus, the present images suggest that Cen~A is unique or at least unusual
in having such large quantities of molecular gas associated with its
shells.

In order to quantify the axis ratios and position angles of the CO
distributions, the
integrated intensity maps were fitted with 
elliptical Gaussians.  
The ellipticity and position angle of the 
fitted Gaussians are given in Table \ref{resultstable}
along with estimates of the maximum detected extent of the CO.

\subsection{CO kinematics}
\label{COkin}

\begin{figure}
\includegraphics[scale=0.8]{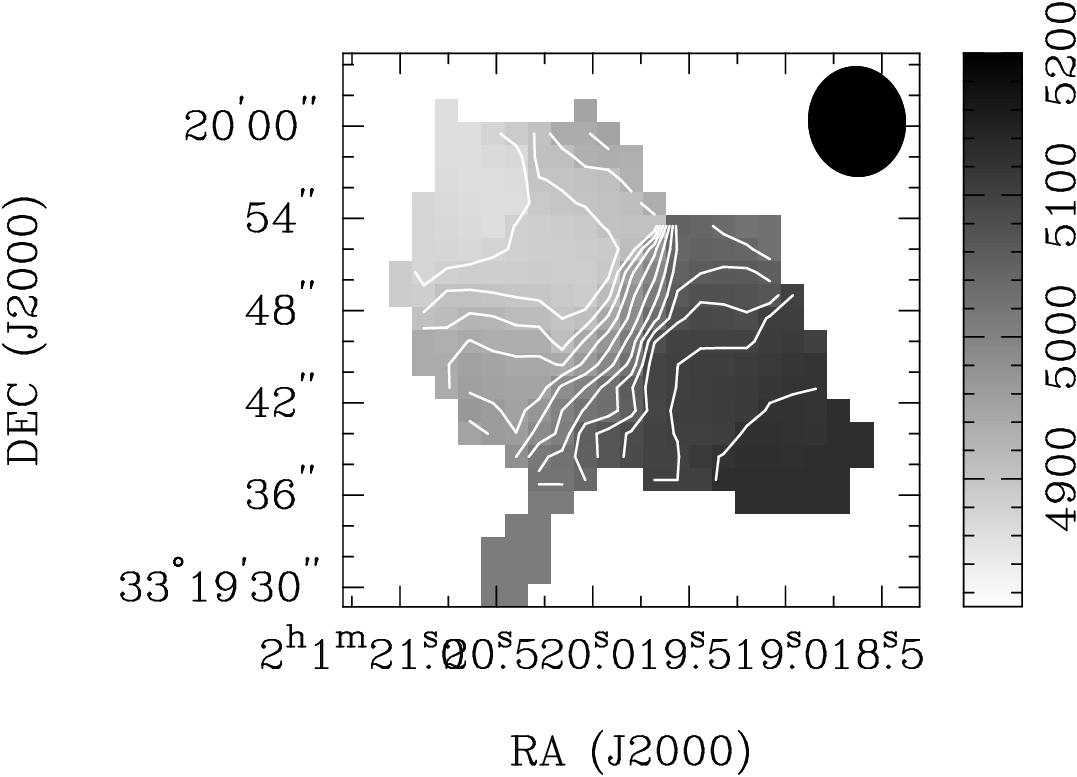}
\caption{UGC 1503 velocity field.  The CO intensity-weighted mean velocity
(moment 1) is shown in grayscale and in contours from 4860 \kms\ to 5120
\kms\ in steps of 20 \kms.  The ellipse shows the beam size.
\label{1503velfield}
}
\end{figure}

\begin{figure}
\includegraphics[scale=0.4,clip,angle=-90]{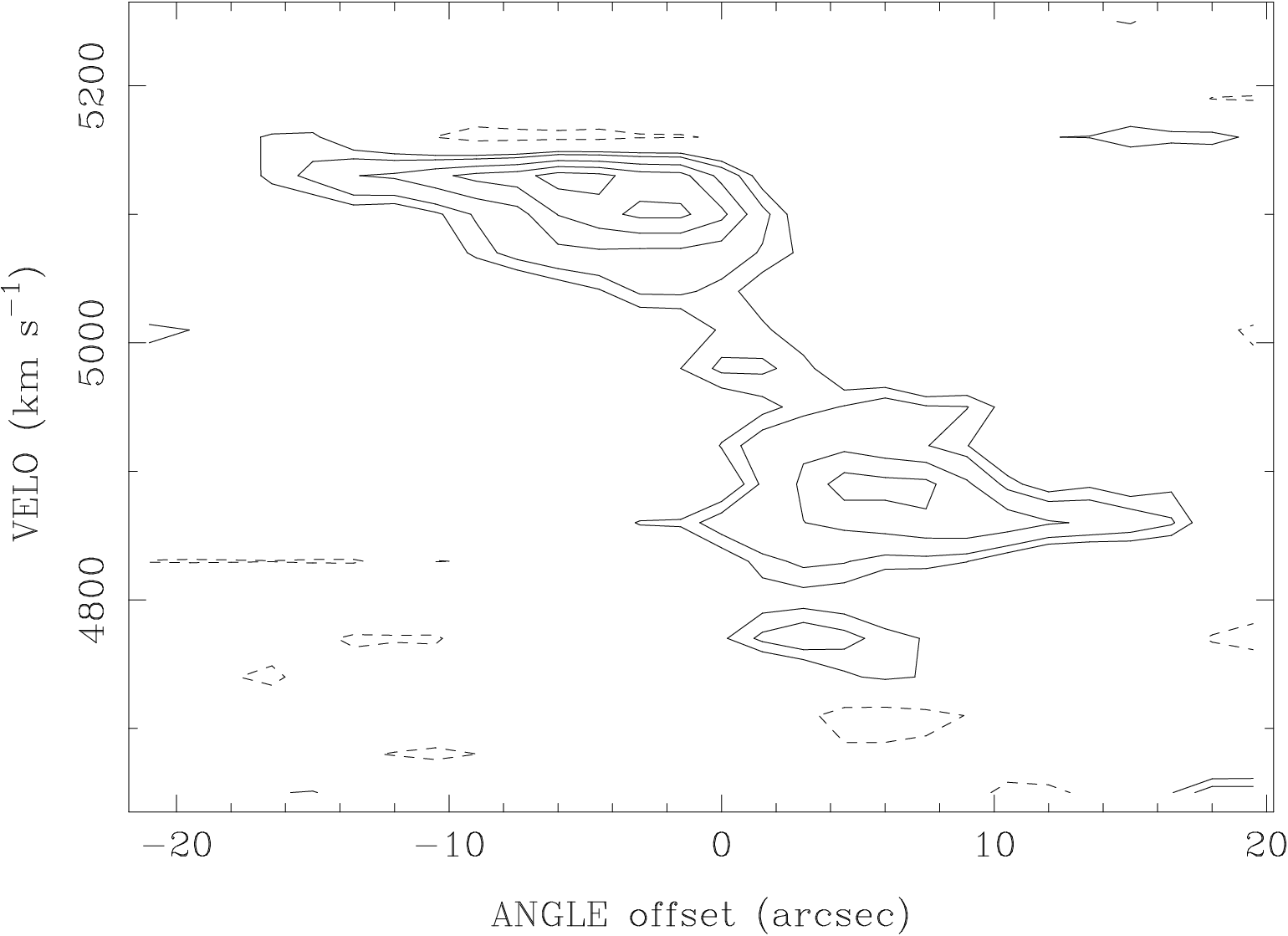}
\caption{UGC 1503 position-velocity diagram.  This slice is centered on the
kinematic center of the molecular gas (RA=02 01 19.8, Dec=+33 19 47, J2000) and
follows the kinematic major axis at $-123$\deg.
Contour levels are $-$20, 20, 30, 50, 70, and 90 percent of 61.1 \mjb.
\label{1503pv}
}
\end{figure}

\begin{figure}
\includegraphics[scale=0.8]{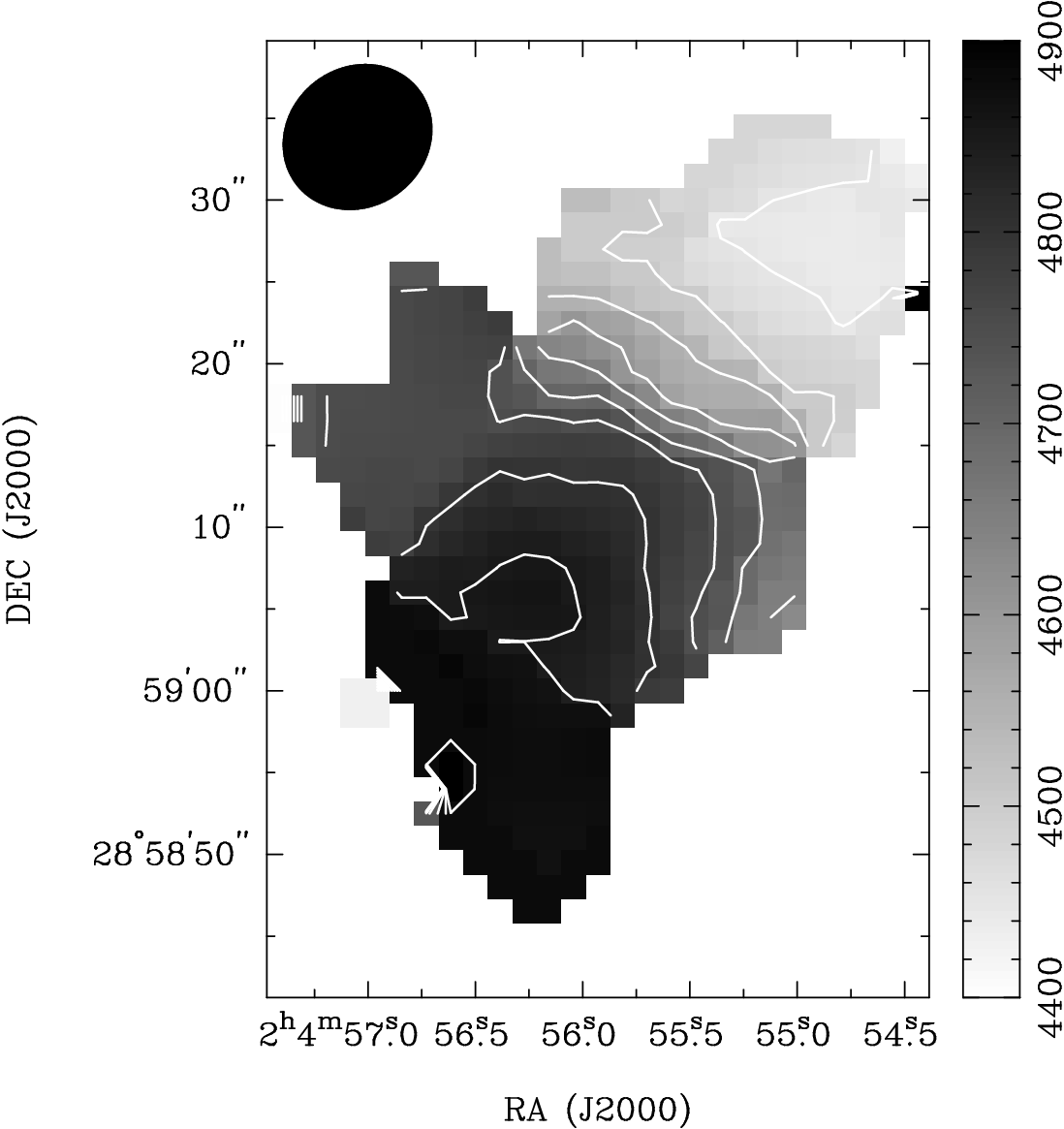}
\caption{NGC 807 velocity field.  The CO intensity-weighted mean velocity
(moment 1) is shown in grayscale and in contours from 4450 \kms\ to 4900
\kms\ in steps of 50 \kms.  The ellipse shows the beam size.
\label{807velfield}
}
\end{figure}

\begin{figure}
\includegraphics[scale=0.4,clip,angle=-90]{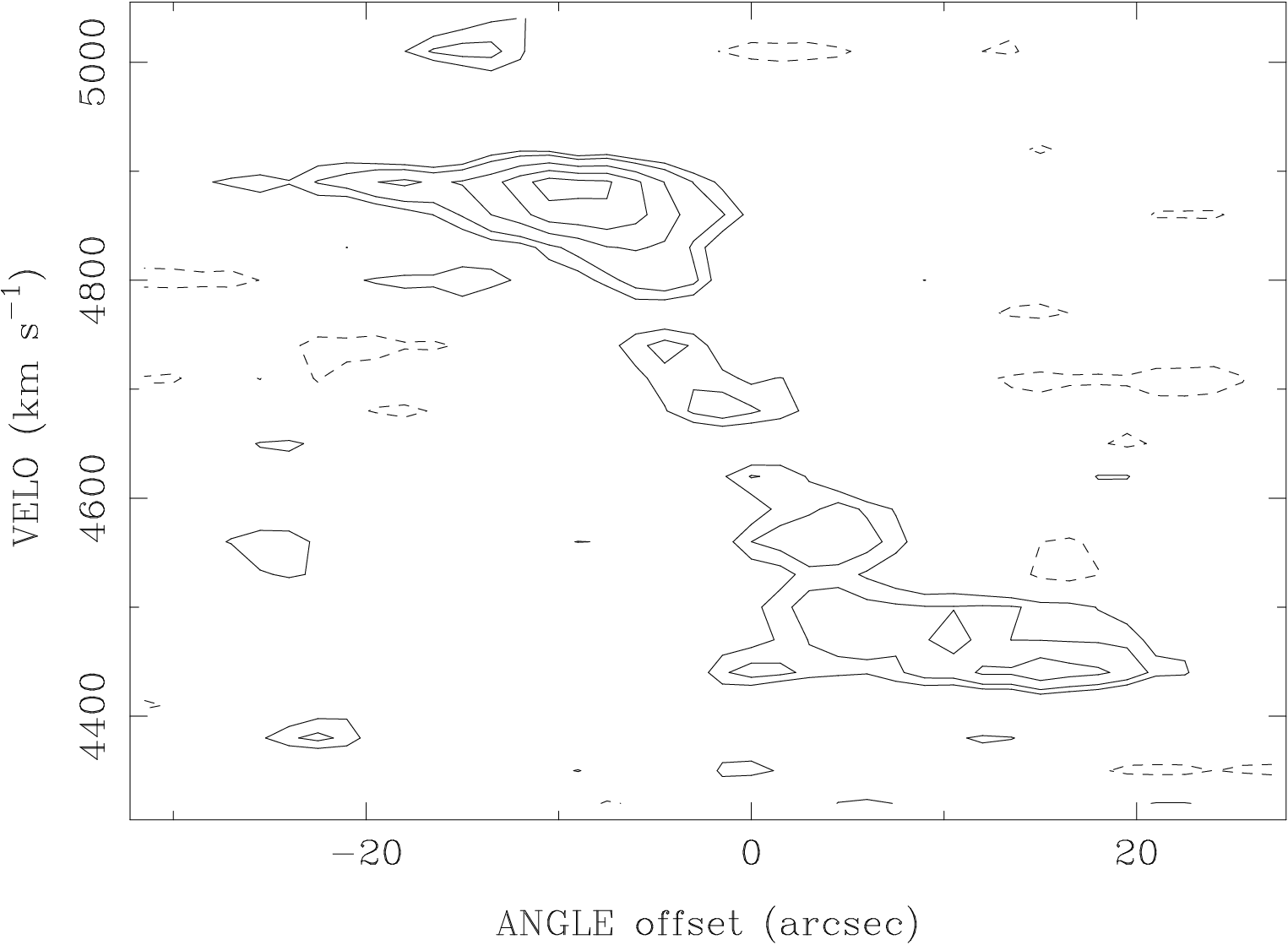}
\caption{NGC 807 position-velocity diagram.  This slice is centered on the
kinematic center of the molecular gas (RA=02 04 55.6, Dec=+28 59 17, J2000) and
follows the kinematic major axis at 149\deg.
Contour levels are $-$20, 20, 30, 50, 70, and 90 percent of 56.8 \mjb.
\label{807pv}
}
\end{figure}

\begin{figure}
\includegraphics[scale=0.8]{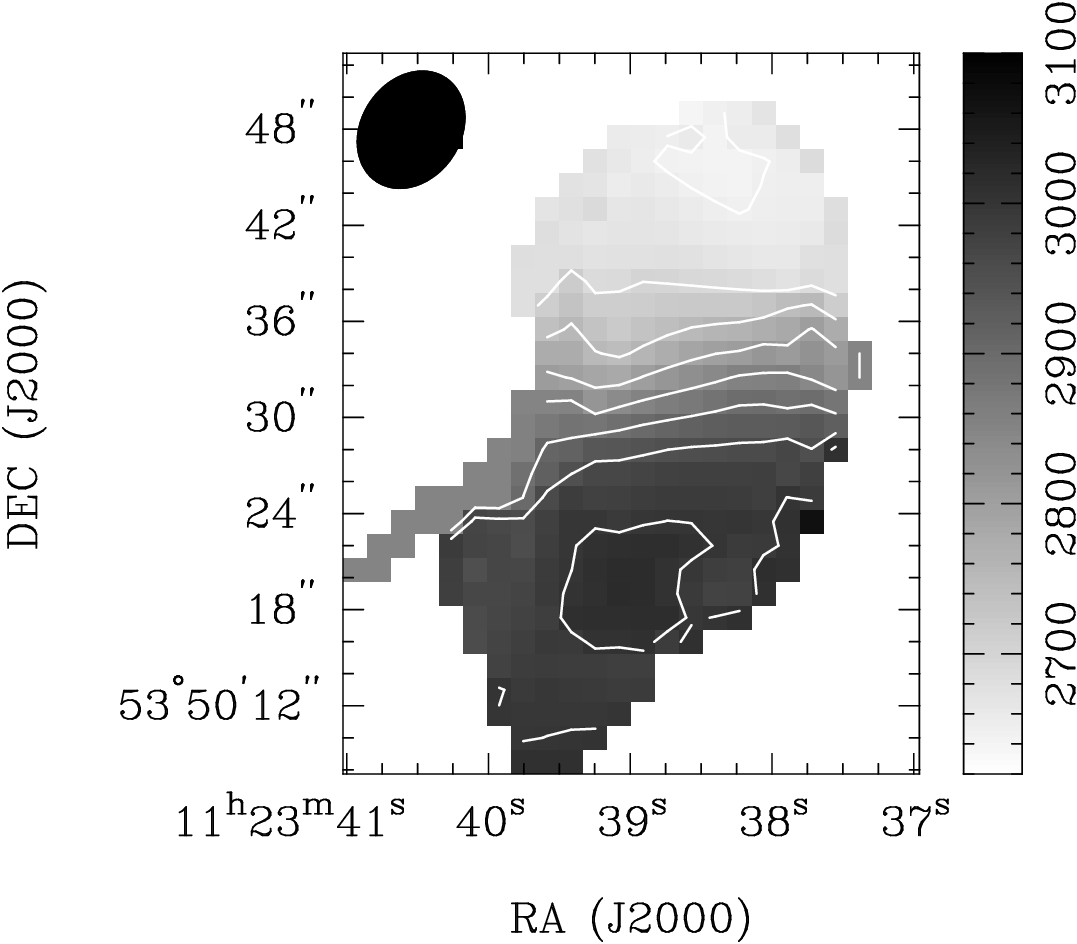}
\caption{NGC 3656 velocity field.  The CO intensity-weighted mean velocity
(moment 1) is shown in grayscale and in contours from 2650 \kms\ to 3000
\kms\ in steps of 50 \kms.  The ellipse shows the beam size.
\label{3656velfield}
}
\end{figure}

\begin{figure}
\includegraphics[scale=0.4,clip,angle=-90]{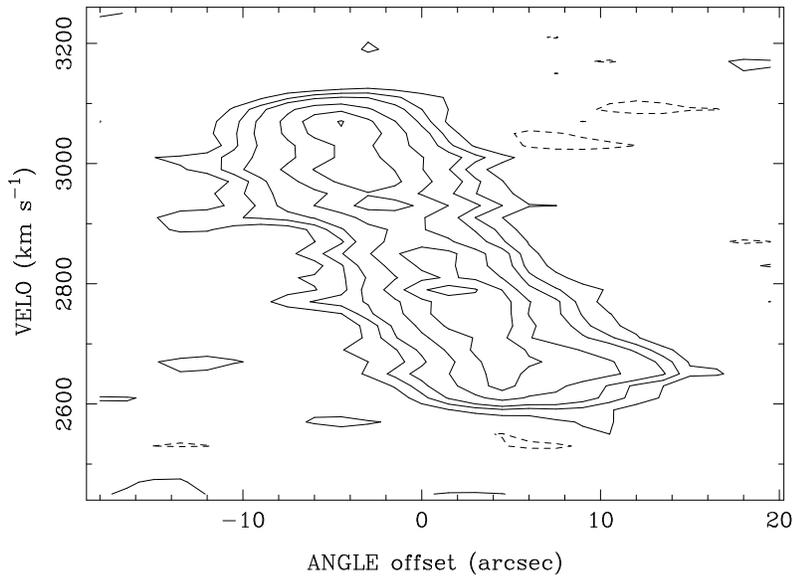}
\caption{NGC 3656 position-velocity diagram.  This slice is centered on the
morphological center of the molecular gas (RA=11 23 38.57, Dec=+53 50 32.3,
J2000) and
follows the kinematic major axis at 191\deg.
Contour levels are $-$20, $-10$, 10, 20, 30, 50, 70, and 90 percent of
292.7 \mjb.
\label{3656pv}
}
\end{figure}

\begin{figure}
\includegraphics[scale=0.8]{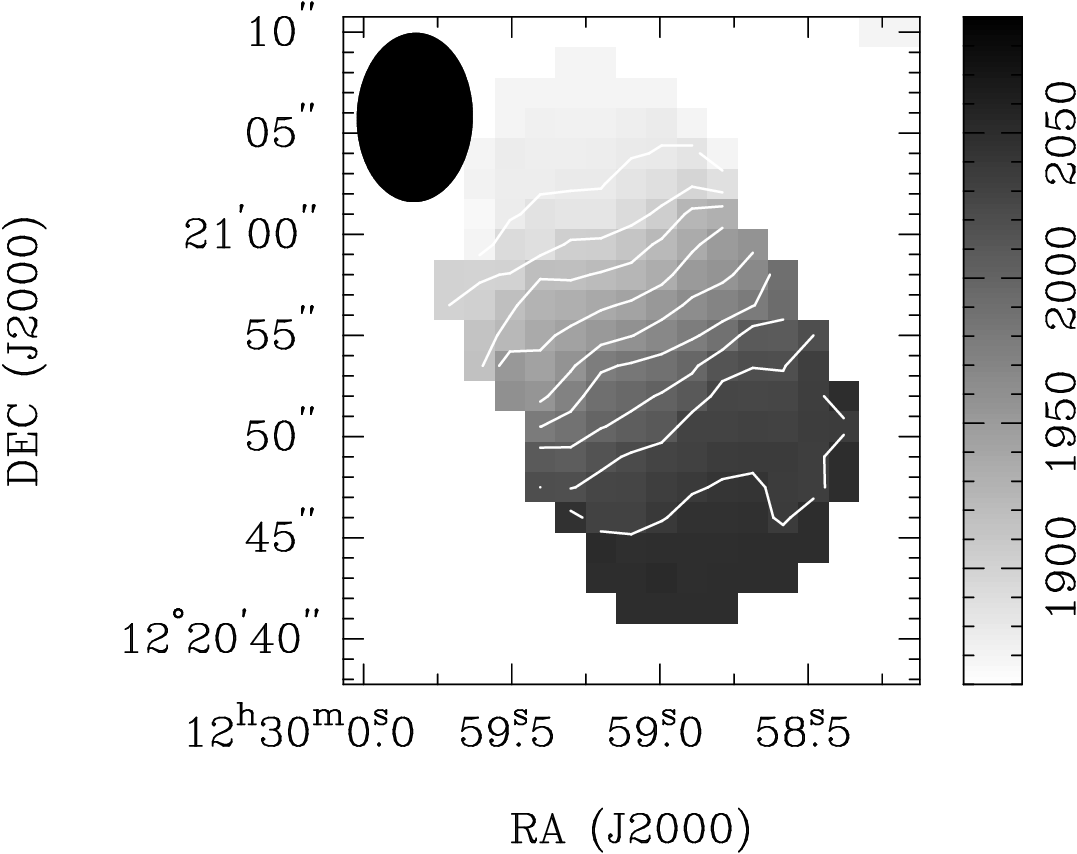}
\caption{NGC 4476 velocity field.  The CO intensity-weighted mean velocity
(moment 1) is shown in grayscale and in contours from 1860 \kms\ to 2040
\kms\ in steps of 20 \kms.  The ellipse shows the beam size.
\label{4476velfield}
}
\end{figure}

\begin{figure}
\includegraphics[scale=0.4,clip,angle=-90]{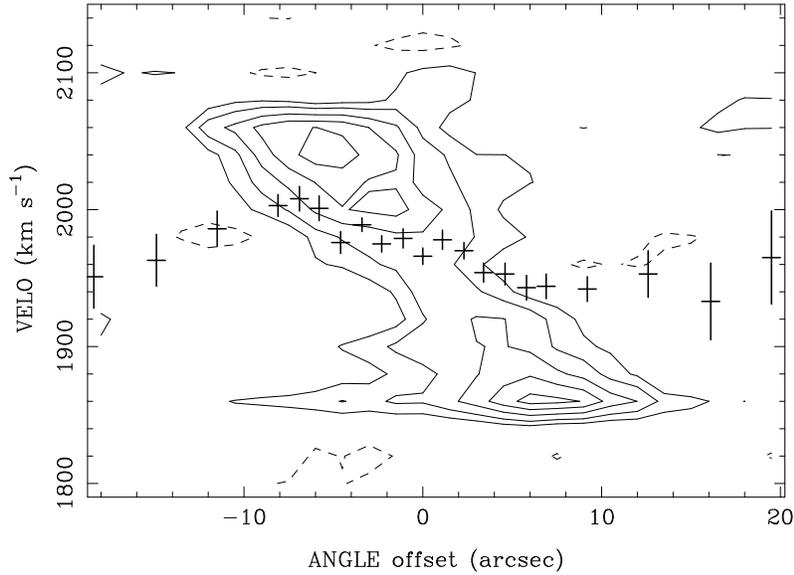}
\caption{NGC 4476 position-velocity diagram.  This slice is centered on the
morphological center of the molecular gas (RA=12 29 59.06, Dec=+12 20 54.2,
J2000) and
follows the kinematic major axis at $-152$\deg.
Contour levels are $-15$, 15, 30, 50, 70, and 90 percent of
125.1 \mjb.
The crosses indicate stellar velocities measured along the major axis by
\citet{simien97}; they have been uniformly shifted in velocity
by 12 \kms\ to make the systemic velocity of the stars agree with that
of the molecular gas.
The difference between heliocentric velocities (in the stellar data) and
LSR (in the CO) is 4 \kms\ at this position.
\label{4476pv}
}
\end{figure}

\begin{figure}
\includegraphics[scale=0.8]{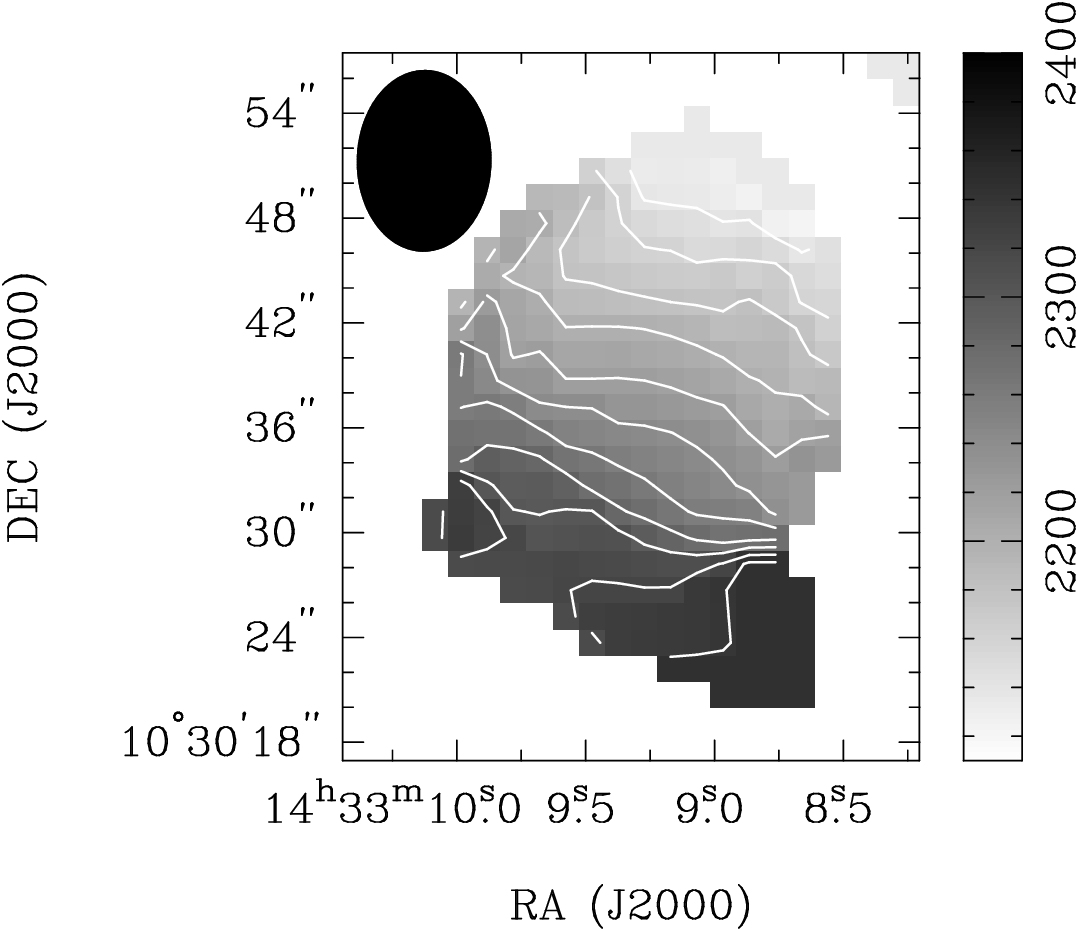}
\caption{NGC 5666 velocity field.  The CO intensity-weighted mean velocity
(moment 1) is shown in grayscale and in contours from 2120 \kms\ to 2340
\kms\ in steps of 20 \kms.  The ellipse shows the beam size.
\label{5666velfield}
}
\end{figure}

\begin{figure}
\includegraphics[scale=0.4,clip,angle=-90]{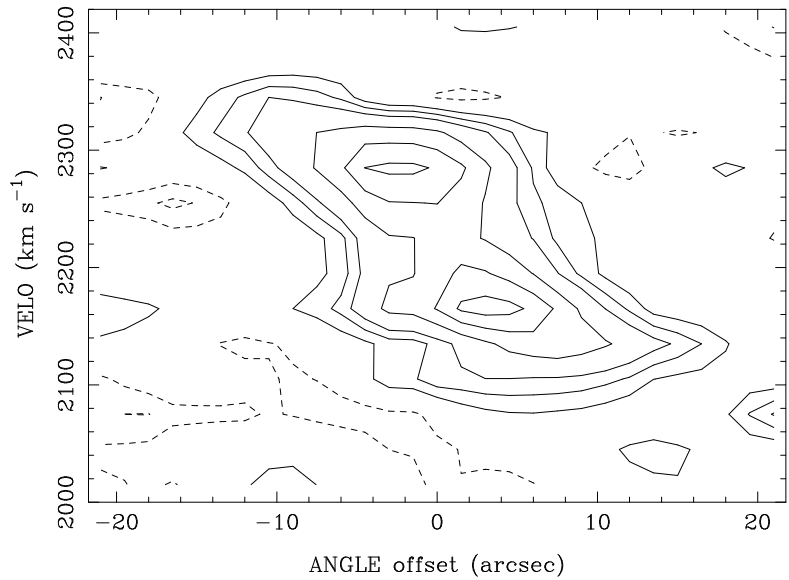}
\caption{NGC 5666 position-velocity diagram.  This slice is centered on the
morphological center of the molecular gas (RA=14 33 09.25, Dec=+10 30 38.7,
J2000) and
follows the kinematic major axis at 165\deg.
Contour levels are $-$20, $-10$, 10, 20, 30, 50, 70, and 90 percent of
149.7 \mjb.
\label{5666pv}
}
\end{figure}

Initial kinematic analysis of the molecular gas was performed by fitting
simple solid body rotation profiles to the velocity fields in Figures
\ref{1503velfield}, \ref{807velfield}, \ref{3656velfield},
\ref{4476velfield}, and \ref{5666velfield}.
NGC~807 and UGC~1503 were also fit with profiles that rise in their inner
parts and then flatten.
None of these fits constrain the disk inclination angles $i$ or maximum
rotation velocity particularly well; the product $V \sin{i}$ is better
constrained.  
But this procedure does give good estimates of the position angle of the kinematic
major axis (Table \ref{resultstable}), which were used for the position-velocity
plots in figures \ref{1503pv}, \ref{807pv}, \ref{3656pv}, \ref{4476pv},
and \ref{5666pv}.  Since the fits are 
weighted by the integrated intensity, the fitted angle is really
the kinematic major axis at {\it small radii}.  The only case in which the
major axis clearly varies with radius is NGC~3656, which is described in
more detail below.

Table \ref{resultstable} compares the morphological 
major axis described in section \ref{COmorph} 
to the kinematic major axis for each galaxy.
In two cases (NGC~5666 and NGC~4476) the two axes are aligned to within
a degree or so, well within the combined errors.
In these galaxies all of the evidence is consistent with the 
molecular gas being in an intrinsically circular disk.

In one case (NGC~3656) there is a dramatic misalignment of nearly 17\deg\
between the two angles in Table \ref{resultstable}, 
much larger than the combined errors of the fits. 
This misalignment is clearly visible in Figure \ref{3656velfield} from the fact that the
kinematic minor axis (the isovelocity curve at the systemic velocity)
is not perpendicular to the kinematic major axis at large radii or to the
morphological major axis.
However, the kinematic major axis at large radii does appear to be closely
aligned with the morphological major axis.
Thus, the molecular gas in NGC~3656 is either in a warped disk or is on
elliptical orbits in a non-axisymmetric potential \citep{binney98}.
Optical images of this galaxy clearly show an S-shaped dust lane (e.g.\
\citet{balcells01}) in which the twist of the dust lane is in exactly
the sense needed to explain the twist in the CO kinematic major axis,
so the CO probably follows the warped dust lane.

The remaining two cases (NGC 807 and UGC 1503) show moderate misalignments
of 6\deg\ and 12\deg\ between the CO morphological and kinematic major
axes.  These
misalignments are nominally greater than the uncertainties in the fits, but
they are probably not reliable.
These are the two galaxies for which the morphological position angles are
the most questionable because the integrated intensity contours are the
least elliptical.

The position-velocity diagrams show steeply rising,
approximately solid body rotation regions in the center of each galaxy.
In NGC~4476 and NGC~5666 the CO does not extend past
the region of solid body rotation, which is at least three to four beams
across.
In NGC~3656 there are some signs that the CO rotation curve
flattens near the edges of the CO distribution;  note in particular
the low-velocity side of the position-velocity diagram (Figure
\ref{3656pv}) and the velocity field (Figure \ref{3656velfield}).
In NGC~807 and UGC~1503 (the most luminous galaxies of the sample, with
the largest CO disk linear sizes) the rotation curve clearly
turns over and becomes flat at radii of approximately 2.0 kpc (NGC~807) and 1.4 kpc
(UGC~1503). 

For these latter two galaxies whose rotation curves turn over, the
kinematic centers of the gas are coincident with the optical centers of the
galaxies \citep{cotton99} within an arcsecond or so (10\% of the 
beam).
For NGC~4476, NGC~3656, and NGC~5666 the kinematic centers are not well constrained, but
the morphological centers of the molecular gas disks are closely coincident
with the optical centers. 

\begin{figure}
\includegraphics[scale=0.4,clip,angle=-90]{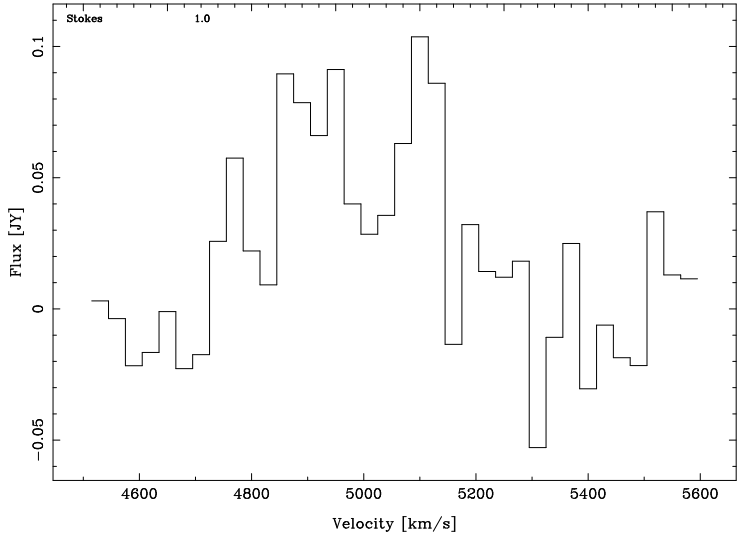}
\caption{CO spectrum of UGC~1503. 
The spectrum was constructed by first using the integrated intensity
image (Figure \ref{1503stars+co}) to define an irregular mask region
within which the emission is located.  The intensity was integrated over
the same spatial region for every channel, so the noise in the line-free
regions of the spectrum should be indicative of the noise on the line as
well.
\label{1503spectrum}
}
\end{figure}

\begin{figure}
\includegraphics[scale=0.4,clip,angle=-90]{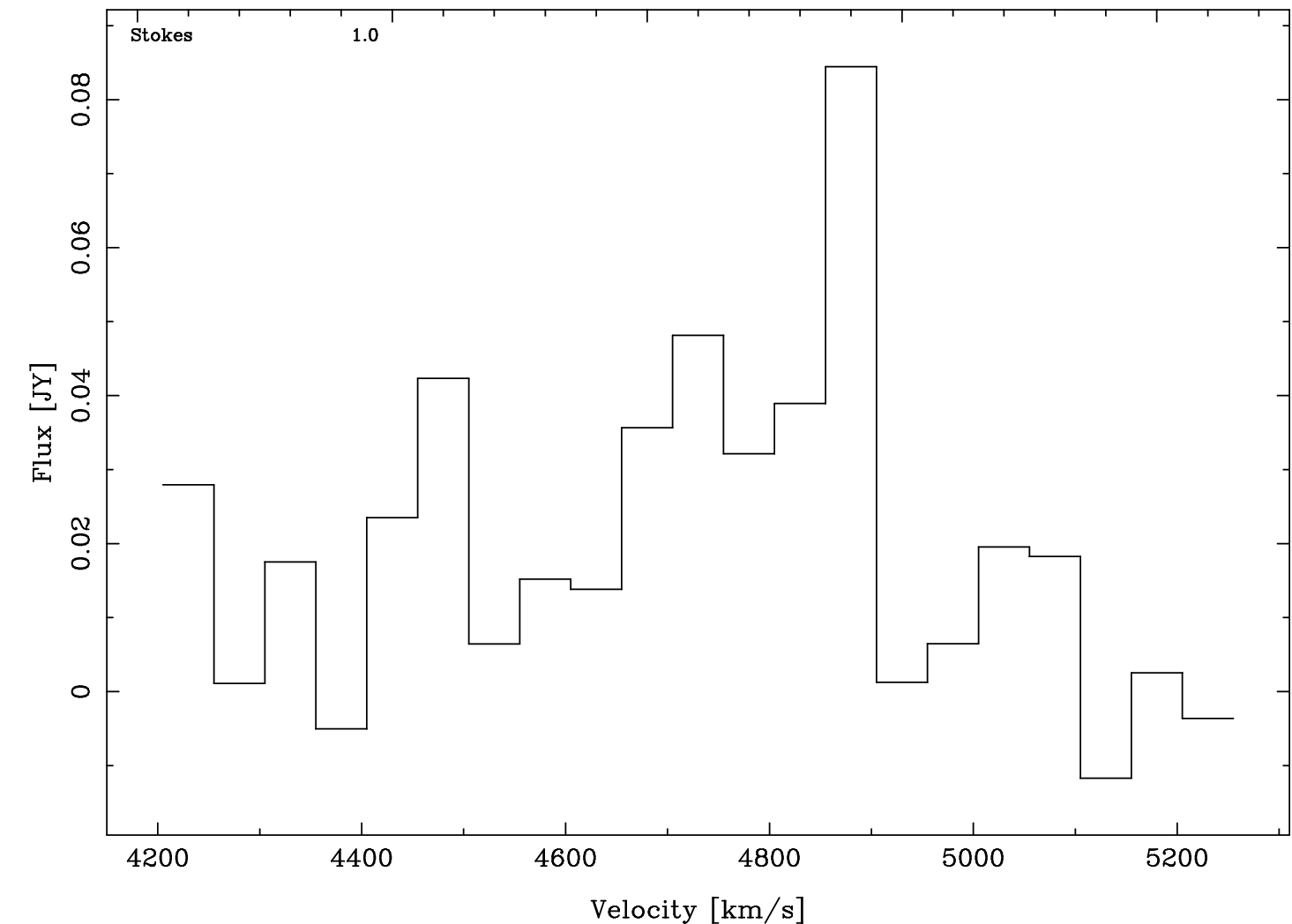}
\caption{CO spectrum of NGC~807, constructed in the same manner as for
Figure \ref{1503spectrum}. 
\label{807spectrum}
}
\end{figure}

\begin{figure}
\includegraphics[scale=0.4,clip,angle=-90]{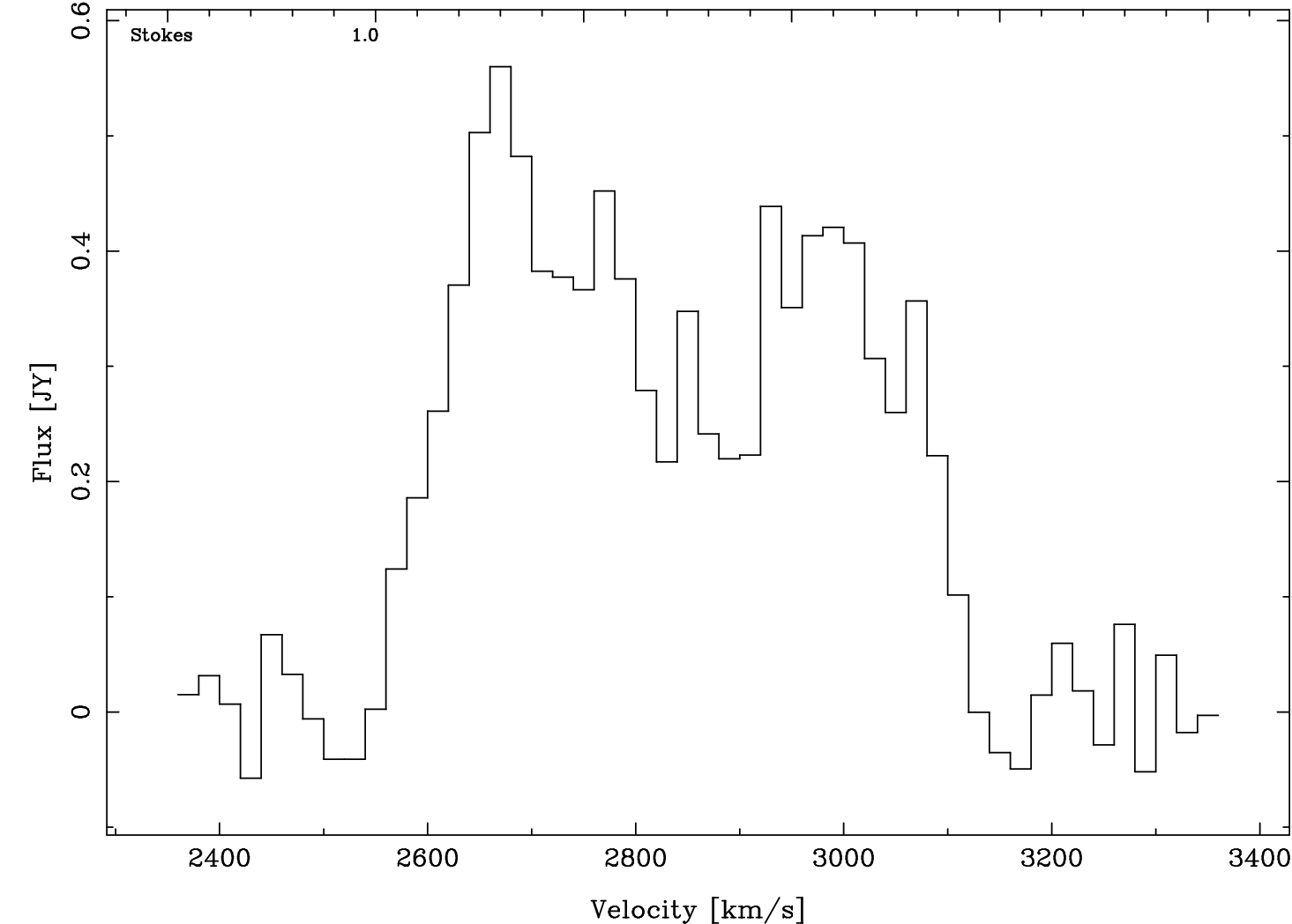}
\caption{CO spectrum of NGC~3656, constructed in the same manner as for
Figure \ref{1503spectrum}.
\label{3656spectrum}
}
\end{figure}

\clearpage

\begin{figure}
\includegraphics[scale=0.4,clip,angle=-90]{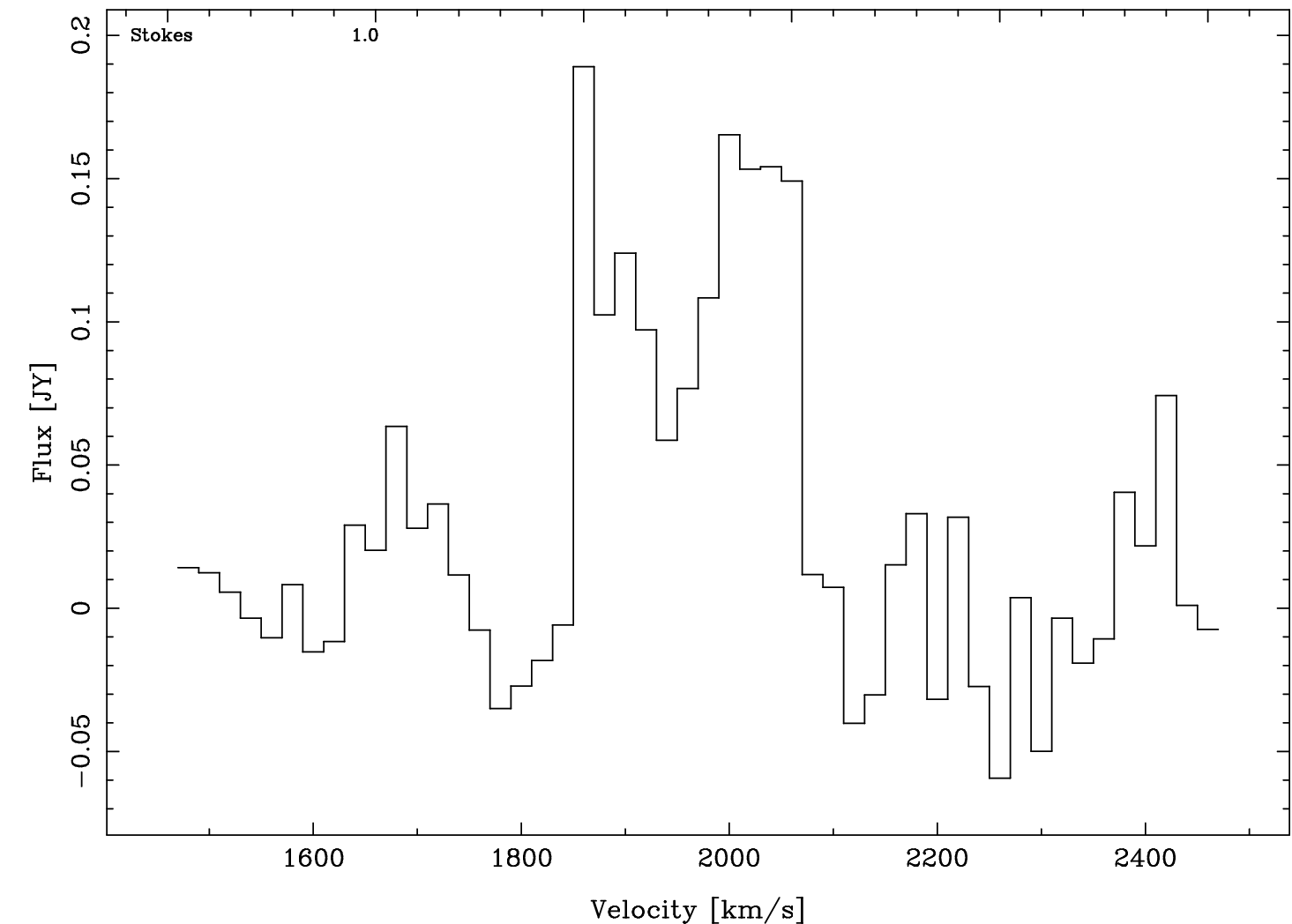}
\caption{CO spectrum of NGC~4476, constructed in the same manner as for
Figure \ref{1503spectrum}.
\label{4476spectrum}
}
\end{figure}

\begin{figure}
\includegraphics[scale=0.4,clip,angle=-90]{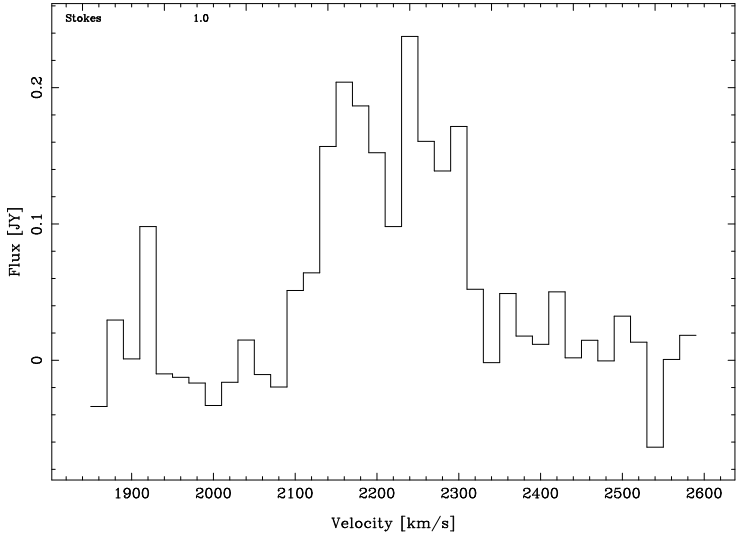}
\caption{CO spectrum of NGC~5666, constructed in the same manner as
Figure \ref{1503spectrum}.
\label{5666spectrum}
}
\end{figure}

\subsection{Dynamical Masses}

These interferometric observations provide two important pieces of
information that are missing from single-dish CO surveys of ellipticals:
the linear sizes and axial ratios of the molecular disks.
If the disks are assumed to be intrinsically circular, with gas on
circular orbits, the inclination angle of the disk is given by
$i \geq \cos^{-1}(b/a)$ 
where $b/a$ is the minor/major axis length ratio and the equality is achieved only in the limit that the disk is very
thin.
The observed gas velocities can then be used to calculate the dynamical
mass interior to the disk's outer edge:
$$M_{dyn} = (2.33\times 10^5\: M_\odot)\:\: V^2\: R$$
where $R$ is the radius of the outer edge in kpc and $V$ is the observed velocity in
\kms, corrected for inclination.
The implied dynamical masses (Table \ref{masstable}) range from a
few $\times$ \tenup{9} \solmass\ to nearly \tenup{11} \solmass\ interior
to the edge of the CO disk,
and the observed masses of molecular gas are a few percent of
these dynamical masses.
Table \ref{masstable} also gives the orbital time for gas at the edges of
the CO disks.

\begin{deluxetable}{lcccccccc}
\tablewidth{0pt}
\tablecaption{Dynamical Masses
\label{masstable}}
\tablehead{
\colhead{Galaxy} & \colhead{$i$} & \colhead{$V \sin{i}$} & \colhead{V} &
\colhead{R} & \colhead{M$_{min}$} & \colhead{M$_{max}$} &
\colhead{M(H$_2$)/M$_{dyn}$} & \colhead{$t_{orb}$} \\
\colhead{} & \colhead{\deg} & \colhead{\kms} & \colhead{\kms} &
\colhead{kpc} & \colhead{\tenup{10} \solmass} & \colhead{\tenup{10} \solmass}
& \colhead{} & \colhead{\tenup{8} yr}
}
\startdata
UGC 1503 &  51 & 130 (10) & 167 (13) & 5.5 (0.6) & 2.2 & 3.6 & 0.050--0.082
& 2.0 \\
NGC 807  &  65 & 230 (10) & 254 (11) & 6.2 (0.6) & 7.6 & 9.3 & 0.015--0.018
& 1.5 \\
NGC 3656 &  74 & 270 (10) & 280 (10) & 3.7 (0.4) & 6.3 & 6.8 & 0.070--0.075
& 0.82 \\
NGC 4476 &  65 & 100 (10) & 110 (11) & 1.2 (0.1) & 0.27 & 0.34 &
0.034--0.042 & 0.66 \\
NGC 5666 &  27 & 100 (20) & 217 (43) & 2.4 (0.2)  & 0.55 & 2.6 &
0.022--0.10  & 0.75 \\
\enddata
\tablecomments{The inclination angle $i$ of the gas disk is given by
$\cos{i} = 1-\epsilon$, with $\epsilon$ given in Table \ref{resultstable}.
If the gas disk is not thin, $i$ is a lower limit, and the true circular
velocity will be somewhere between $V \sin{i}$ and $V$; the enclosed
dynamical mass will be between M$_{min}$ and M$_{max}$.
The range in M(H$_2$)/M$_{dyn}$ comes from the range in the enclosed
dynamical mass, ignoring the uncertainty in the \htooco\ conversion
factor, which is probably at least 50\% to 100\%.
}
\end{deluxetable}

\subsection{CO vs.\ stellar morphology}\label{opticalangle}

Optical images from the red plates of the second generation Digitized Sky
Survey (DSS)
were used in a comparison of CO and stellar morphologies.
After sky subtraction, elliptical isophotes were fit to the optical images.
The ellipticity, position angle, and center of each isophote were allowed to
vary freely.
The isophote fits are generally not good within a semi-major axis of 5$''$,
where the Sky Survey images are overexposed, or beyond $30''$, where the
images are underexposed.
The region from $10''$ to roughly $30''$ is comparable to the size of the
CO disks, and that is the region I focus on here.
A second round of fitting in which the ellipse centers were held fixed
did not significantly change the results over this radius range.
For comparison purposes, elliptical isophote fits were also performed on
the J, H, and K images of NGC~4476 from the Two Micron All Sky Survey
(2MASS) data.
Isophote fits to NGC~3656 are more complicated because of the fine
structure and the prominent dust lane, so the results of \citet{balcells97}
are used for that galaxy.

Table \ref{angletable} gives the radius range over which the isophote fits are considered
reliable, the mean ellipticity and position angle for each galaxy,
and the dispersion about the mean for roughly 12 independently fitted
annuli in that radius range.
NGC~807, NGC~4476, and NGC~3656 are better described by $r^{1/4}$ 
surface brightness profiles than by exponential profiles over the radius
range in question.
UGC~1503 and NGC~5666 are about equally well fit by either, though the
reliable radius range for NGC~5666 is rather small.
There is no evidence of position angle twist in any of the galaxies
except for NGC~5666, which shows a 10\deg\ change in position angle at 
semi-major axes around 15$''$.
UGC~1503 shows no significant trend in ellipticity with radius, but the 
others do, and the dispersions about the mean ellipticies are determined
mostly by the magnitude of those trends.
The Sky Survey data for NGC~4476 give results which are consistent with
those from the 2MASS data and the work of \citet{simien97};
my fits for UGC~1503 are consistent with those of \citet{fasano89}.

The stellar isophotes are significantly rounder than CO isophotes,
as one would expect in the case where the stars are dynamically hot and gas
is dynamically cold.
The only exception to this statement is NGC~5666, where the stars and the
gas have equal ellipticities within the errors.
Table \ref{angletable} also gives the misalignment angle between the optical major axis
and the CO kinematic axis;
except for NGC~3656, which is close to a minor axis gas/dust disk,
they are close to zero.
In other words, four of the five galaxies have remarkably well-aligned (within 13\deg)
major axis gas disks.

\begin{deluxetable}{lcrrc}
\tablewidth{0pt}
\tablecaption{Alignment with Optical Major Axis
\label{angletable}}
\tablehead{
\colhead{Galaxy} & \colhead{Radii} & \multicolumn{2}{c}{Optical Morphology} & 
\colhead{Misalignment} \\
\colhead{} & \colhead{$''$} & \colhead{$\epsilon$} & \colhead{PA ($\deg$)} & 
\colhead{$\rm |\Delta PA|$ (\deg)} 
}
\startdata
UGC 1503 & 10--27 & 0.21 (0.02) & $-$124 (2) & 1 (3) \\
NGC 807  & 10--30 & 0.35 (0.05) & 140 (1) & 9 (3) \\
NGC 3656 & \nodata & 0.20 (0.04) & 110 (5) & 81 (6) \\
NGC 4476 & 10--30 & 0.35 (0.02) & $-$155 (1) & 3 (2) \\
NGC 5666 & 10--20 & 0.14 (0.02) & 151 (4) & 13 (5) \\
\enddata
\tablecomments{Optical ellipticities and position angles are mean values
over the semimajor axis range indicated under ``Radii.''
Optical morphology information for NGC~3656 is taken from Balcells
(1997) so is not restricted by the overexposed/underexposed reigons of
the DSS.
The parenthesized values in the $\epsilon$ and PA columns are the
dispersion about the mean $\epsilon$ and PA over that radius range.  
Because these dispersions are normally larger than the
formal fit uncertainties, they indicate something about the 
magnitude of possible
radial variations in $\epsilon$ and PA.
The misalignment angle $\rm \Delta PA$ is the difference between the 
molecular gas's kinematic PA (Table \ref{resultstable}) and the optical major axis PA.
}
\end{deluxetable}

\section{Discussion}

\subsection{The origin of the molecular gas}\label{origin}

Two ideas about the origin of molecular gas in ellipticals are
(1) that the gas came from mass loss from the galaxy's own evolved stars
or (2) that the gas was acquired in an interaction or a merger with
another gas-rich galaxy.
In the second category I also include the idea that the molecular gas in
ellipticals may simply be leftover from the formation of the elliptical,
if ellipticals are formed by the merger of two roughly equal-mass spiral
galaxies \citep{toomre72}.
In the first model, internal stellar mass loss,
\citet{faber76} estimate that mass loss rates would be 1.5
\solmass~yr$^{-1}$ per \tenup{11} L$_\odot$ of optical luminosity.
Over \tenup{10} years this gas would be comparable to the observed
gas masses, at least within a factors of a few (the uncertainties in the
\htooco\ conversion factor are at least factors of a few).
The difficulty with this model is that the cold gas
contents of elliptical galaxies are uncorrelated with their optical
luminosities \citep{knapp85, lees91}.
Furthermore, the stellar mass loss
is thought to be shock-heated by the stellar velocity dispersions 
to X-ray temperatures, and the hot plasma is thought to destroy
dust grains on relatively short timescales \citep{wiklind01}.

The second idea, that the molecular gas in ellipticals has been 
acquired in a major or minor merger, may plausibly agree with the 
data presented here.
\citet{barnes02} has shown, {\it via} numerical simulations,
that the merger of two gas-rich
spiral galaxies produces systems which are qualitatively similar to 
the present
sample of ellipticals and their gas disks.
Some of the gas loses its angular momentum in shocks and falls to the 
nucleus of the galaxy, but
up to 60\% of the original gas contained in the spirals can form
a rotationally supported gas disk with a radius of up to 20 kpc.
Thus, the gas disks formed in these simulations are large enough
to explain the observed gas disks
(keeping in mind that the 
mass and size of the simulated disks are highly dependent on the geometry
of the interaction).
More careful analysis of the distribution and kinematics of the molecular
gas in these ellipticals offers some important insight into these
competing models and the origin of the molecular gas.

\subsubsection{Gas and stellar kinematics}

Figure \ref{4476pv} shows CO and stellar kinematics along the major axis
of NGC~4476.
The stellar kinematics are taken from \citet{simien97}, who also
give a value of 14.8\asec\ for the effective radius (\re) of this galaxy.
The CO velocity rises linearly with radius to a maximum velocity of 100
\kms\ at 7\asec\ radius.
The stellar velocities also rise linearly with radius to a maximum
rotation velocity of 35 \kms\ (3 times smaller than the CO velocity) at 7\asec\ radius.  Beyond 7\asec,
the stellar velocities appear to decline again so that there is little
sign of rotation at \re.

The stellar velocity dispersion is 65\error 13 \kms\ in the center of the
galaxy, much larger than the stellar rotation velocities; the stars are 
primarily pressure-supported rather than
rotationally supported.  This means that the {\it stellar} rotation 
velocities in Figure
\ref{4476pv} significantly underestimate the circular rotation speed
of the galaxy (the asymmetric drift effect).   The cold molecular gas probably
gives a good indication of the circular speed.
But the stellar rotation velocities {\it do} indicate the specific
angular momentum of the stars.
Within \re, the specific angular momentum of the gas is about three
times larger than that of the stars.
Furthermore, if the observed trend in stellar rotation continues beyond 
\re, then the stars in the outer parts of the galaxy also have very
small specific angular momentum.
If internal stellar mass loss had produced this molecular gas, one would expect
the gas to have the same specific
angular momentum as the stars, or perhaps smaller, and this is clearly not the case.
The only way to  reconcile the specific angular momenta of the gas and
the stars is to suppose that they have very different inclinations to the
line of sight, but that would again be unlikely if the gas originated in
the stars.
An external origin for the gas in NGC~4476 is strongly favored.

A similar situation seems to be true for NGC~3656.
\citet{balcells90} obtained stellar kinematics along two
position angles, both of which are 50\deg\ away from the CO major axis.
They infer that the maximum stellar rotation velocity in the inner
10\asec\ is on the order of 50 \kms\ and that the 
stellar rotation axis is very close to what is now known to be the
CO rotation axis.
The CO rotation velocity is 270 \kms, five times larger than that of the
stars, but firm statements about the specific angular momenta of the
stars and the gas are not possible in this case because the stellar
rotation curve may still be rising at large radii.
Similar comparisons of gas and stellar kinematics for the other galaxies
in the sample will be vital for a broader understanding of the origin of
the molecular gas
in ellipticals.

\subsubsection{Orientation of molecular gas and dust}

There are two cases in the present sample for which published 
optical images clearly show that there is a very close correspondence between
the dust and molecular gas distributions.
These cases include NGC~3656, mentioned in section \ref{COkin}, and
NGC~4476. 
\citet{tomita00} show that the dust in NGC~4476 is settled into a very
regular, highly inclined disk of diameter 20\asec; the CO disk is also
very regular, highly inclined, and has diameter
27\asec.
The other galaxies of the present sample do not have good dust images
in the literature and dust features are not visible in the Digitized Sky
Survey images, so detailed comparisons of their dust {\it vs.} CO
morphologies will require higher quality optical images.

As mentioned in section \ref{origin}, the cold gas contents of ellipticals
are unrelated to their optical luminosities, and this fact is usually
interpreted as evidence of the gas's external origin \citep{wardle86,
lees91}.   But the cold gas and dust {\it distribution}
within an individual galaxy are clearly not independent of the stars.
\citet{vandokkum95}
studied the orientations of dust features in
HST images of ellipticals; they found that dust features with 
semimajor axes smaller than 250 pc are well aligned with the optical
major axes of their host galaxies.
More recent workers \citep{verdoes99, martel00, tran01}
classify dust features into two classes: (1) smooth, regular
disks and (2) irregular lanes or filaments.  They find that the disks are 
closely aligned with their host galaxies' major axes whereas the lanes
are randomly oriented. 
The interpretation which is common to all of these studies is that the
dust has been acquired from an external source;
the initial orientation of the dust features is random and their
structure is irregular,
but the dust gradually settles into the preferred plane of the galaxy
and becomes a regular disk.

The close alignments between the CO disks and the optical major axes of
the present sample are consistent with what is seen in the dust studies
mentioned above, if the present sample of CO disks correspond to the
older and more relaxed dust systems.
Note, however, the curious fact (probably a selection effect) that the
CO disks studied here are larger than the dust features seen
by \citet{vandokkum95}, \citet{verdoes99}, and \citet{tomita00}.  Most of the
dust features have diameters smaller than 2 kpc whereas only one of the
CO disks (NGC~4476) is that small.

\subsection{The shape of these galaxies}

Many attempts have been made to infer the intrinsic shape distribution of
elliptical galaxies from their optical photometry and kinematics, 
but the true shape distribution is still poorly known because it is 
model-dependent.  
Some ellipticals may be oblate, but it seems highly unlikely that {\it
all} of them are (\citet{khairulalam02, bak00} and references therein).
However, the preponderance of major axis disks in the present sample 
suggests that the majority of the sample galaxies are oblate.
The principal plane of an oblate spheroid is perpendicular to the
short axis, and this short axis always projects onto the apparent minor
axis if the galaxy is axisymmetric \citep{dezeeuw89}.
Thus, a relaxed gas disk in an oblate galaxy should be aligned with the 
optical major axis.

At the present time it is not clear whether there is a discrepancy
between the number of oblate ellipticals in the present sample and in the
optical studies mentioned above.
The number of elliptical galaxies with CO maps is still too small to 
confirm or reject the hypothesis that the CO-sample has been drawn from
the same parent population as the optical studies.
But when the number of ellipticals with CO maps is significantly greater,
it should prove interesting to investigate whether the amount of
molecular gas in these galaxies is correlated with their intrinsic shape.

\subsection{Star formation and the future of the molecular gas}

Molecular gas is understood to be the raw material for star formation;
the transformation of gas into stars will create rotationally supported
stellar disks within these ellipticals.
An estimate of the masses of the stellar disks can be obtained from
Table \ref{masstable}, which shows that the molecular gas masses in the sample
galaxies are a few percent of the dynamical masses within the edge of the
CO disks.  Comparing the masses of the
stellar disks to the masses of the spheroidal stellar components
depends on an assumption that the dynamical mass within the CO disk
arises mostly from stars.
This assumption is reasonable for the galaxy
interiors.\footnote{For NGC~4476, the effective radius 14.8\asec\ quoted by
\citet{simien97} is similar to the 13\asec\ maximum radius of the CO
(Table \ref{resultstable}); the CO disk extends to approximately $r_e$.  
The other sample galaxies do not have
published effective radii, but from Figures \ref{1503stars+co},
\ref{807stars+co}, \ref{3656stars+co}, and \ref{5666stars+co} it seems
unlikely that the CO extends to radii much larger than $r_e$.}
It is also likely that not quite all of the gas will be transformed
into stars.
These assumptions imply 
that the stellar disks will have masses on the
order of a percent of the total stellar mass in the galaxies-- perhaps
somewhat more,
if some of the molecular gas has already been transformed into stars.

The stellar disks which are likely to form out of these molecular disks
will be very similar to the stellar disks which are now known to be
common at least in disky ellipticals.
\citet{scorza98, scorza95, cinzano94},
and others who have done detailed photometric and
kinematic studies of ellipticals find that many ellipticals contain both
the usual spheroidal component (a bulge)
and a stellar disk.  In this respect the ellipticals have structure which
is qualitatively similar to spirals, but with much 
larger bulge/disk ratios \citep{kormendy96}.
\citet{scorza98} and \citet{scorza95} found that the 
stellar disks inside ellipticals are rotationally supported; their 
sizes vary widely but are commonly on the order of \re; and their
disk/bulge (luminosity) ratios are commonly a few percent up to 0.3.
Presumably, smaller stellar disks may exist as well but are more
difficult to detect.
In short, after star formation ceases and the molecular gas is gone, 
the current sample of ellipticals will look much like known disky
ellipticals.

The formation of a rotationally-supported stellar disk 
may already have happened in NGC~4476, where the stellar
rotation appears to die out at the edge of the CO disk.
I propose that careful kinematic analysis of that galaxy will show a small stellar
disk of radius 15\asec\ and circular rotation speed \approx\ 100 \kms\ 
superposed on a largely non-rotating spheroidal population.

Interestingly, \citet{naab01} already predicted the existence of
molecular gas disks in some ellipticals, at least in the scenario where
ellipticals are formed by the merger of two similar-mass disk galaxies.
They found that purely collisionless mergers of disk galaxies do not
reproduce the detailed kinematics of real elliptical galaxies (see also
\citet{bendo00}).
In particular, the collisonless mergers do not have broad retrograde
and steep prograde wings in the stellar line-of-sight velocity
distributions, and those are the signatures of rotationally supported stellar disks.
Therefore, \citet{naab01} suggested that real ellipticals with 
stellar disks must have {\it first} contained gas disks similar to the
ones which are shown here; the gas disks later turned into stellar disks.

\subsection{Stability of the Asymmetries in NGC 807}

The strong asymmetry in the CO distribution of NGC~807 (Section
\ref{COmorph}) is striking,
particularly in view of the fact that the gas should be sheared by
differential rotation on very short timescales.
Figure \ref{807pv} shows that the rotation curve for NGC~807 becomes flat
at a distance of about 6\asec\ (2.0 kpc) from the kinematic center of the
galaxy.  The peak of the CO distribution is about 9\asec\ (2.8 kpc)
from the center, well within the differentially rotating part of the galaxy.
The orbital timescale at the farthest edge of the CO disk in NGC~807 is
1.5\e{8} yr (Table \ref{masstable}), and the orbital timescale at the
turnover point is only 5\e{7} yr.
These numbers are not strongly dependent on the assumed inclination, as
the $\sin{i}$ correction is only 10\% for the CO in NGC~807 (Table \ref{masstable}).

Thus, the effects of differential rotation should have been severe for this gas
over the last \tenup{8} or few $\times$ \tenup{8} years.  It is
difficult to understand how the strong asymmetry (twice as much emission
from the southern half of the galaxy as from the northern half) could be
maintained for as long as \tenup{9} years. 
One possible solution to the problem is that the 
gas was acquired less than \tenup{9} years ago.
On the other hand, an interaction with another galaxy less than \tenup{9}
years ago should leave other traces as well-- features like shells, ripples,
or tails in the optical or HI images, but none have been seen 
\citep{oosterloo99, dressel87}.

There are several other possible solutions to this short timescale for
shearing.  For example, the molecular gas may be much farther from the
center than it appears.  The orbital timescale would then be longer 
than 1.5\e{8} yr.
This scenario implies a highly nonuniform distribution of molecular gas,
which again might be evidence that the gas was acquired from some outside
source.
The molecular gas could be on elliptical orbits; the gas distribution
should then show a peak in the place where the orbital speed is small.
It is also possible that the peak in the CO intensity reflects not a
peak in the total gas density but rather a change in the phase from
atomic to molecular; comparisons with the atomic gas distribution could
address this latter possibility.

One final way to avoid the shearing problem is if the molecular gas
which peaks about 9\asec\ south of the galaxy center (Figures
\ref{807stars+co} and \ref{807pv}) is strongly
self-gravitating.
This feature has a CO flux of 9.2 \jykms, so its 
mass (not including helium) is approximately 4.5\e{8} \solmass.
The mass and the fact that it is not well resolved in the spatial dimension
are consistent with \h2\ densities in the range \tenup{2} to \tenup{3}
\percc\ (4\e{-22} to 4\e{-21} g~\percc, including helium), which are 
typical values for Galactic giant molecular clouds.
From the rotation velocity of the CO we infer that the galaxy itself has
a mass of 4.1\e{10} \solmass\ within 2.8 kpc of the center, giving an average
density of 3.0\e{-23} g~\percc.
In these conditions the Roche limit for molecular gas of density \tenup{2}
to \tenup{3} \percc\ is 1.3 to 2.8 kpc, which implies that the molecular
gas is quite close to the borderline between tidal instability and
stability.
Again, the molecular gas may be farther from the center than it appears,
which would tend to increase its stability.

\subsection{A cautionary note}\label{caution}

Given that there are still only a very small number of ellipticals whose CO
distribution has been mapped, 
it is important to remember that the CO distributions in the present sample
may not be representative of ellipticals in general.
For example, the galaxies have been selected from single-dish surveys
which usually only made one pointing towards the center of each galaxy.
Any ellipticals in which molecular gas avoids the inner 10\asec\ (in radius)
would not have been selected.
Furthermore, the galaxies are selected to be bright at 100\micron, but it
is well known that IRAS is much more sensitive to warm dust than to cold
dust.  Ellipticals whose dust is primarily cold would also not have
been selected.
If interactions increase the FIR luminosity of a galaxy,
the present sample may be biased towards galaxies that have recently
undergone an interaction.
Future progress in understanding the evolution of ellipticals requires
some exploration of these kinds of selection effects.

\section{Summary}

Six elliptical galaxies from the CO survey of \citet{wiklind95}
and one from \citet{sage89} were observed with the BIMA 
and OVRO millimeter arrays at about 8\asec $\times$ 6\asec\ resolution
(0.7 to 2 kpc).
Five of the seven (UGC~1503, NGC~807, NGC~3656, NGC~4476, and NGC~5666)
were detected and their CO emission is resolved
into very regular rotating disks.  
The disk radii are 1 to 6 kpc and rotation velocities are 100 to 280
\kms; orbital timescales at the edge of the disks are around \tenup{8} yr.

Two of the five detected galaxies have CO rotation curves which rise
linearly and then flatten at radii of 1--2 kpc.
Two other galaxies' rotation curves keep rising to the edge of the
CO disk; one may flatten just at the outer edge of the CO.
The sizes, observed velocities, and inclination angles of the disks
enable robust dynamical mass estimates which range from 3\e{9} \solmass\
for NGC~4476 to \tenup{11} \solmass\ for NGC~807.  Of course, the
dynamical mass estimates pertain to the mass interior to the edge of the
CO disk, which for NGC~4476 is about one effective radius.
The H$_2$ masses are only a few percent of the dynamical masses.

Four of the CO disks are aligned within 13\deg\ of their host galaxies' 
optical major axes. 
The high proportion of major axis gas disks suggests that these
ellipticals are oblate.
The exception, NGC~3656 (a merger remnant), has a gas and dust disk 
nearly aligned with its minor axis.
The kinematics of the gas in NGC~3656 show clear evidence for a warp, but
the others do not appear to be warped at the current resolution and
sensitivity.

In one case, NGC~4476, major axis stellar kinematics are available from
the literature.
The stars show some rotation over the radial range where the CO exists,
but outside that range the stellar rotation velocity appears to drop to zero.
The gas has a factor of three or so larger specific angular momentum than
the stars, which strongly suggests that the gas in NGC~4476 has come from an
external source rather than from internal stellar mass loss.
However, the present analysis offers no insight into
the question of whether the gas in NGC~4476 is leftover from a major
merger or was acquired in a minor merger/interaction.
The CO disks are well within the size range of the gas disks which are
created in spiral-spiral merger simulations \citep{barnes02}, so the
major merger hypothesis is plausible in this respect.  Only
one of the galaxies shows clear evidence of a major merger (NGC~3656;
\citet{balcells97, balcells01}).

If the molecular gas in these galaxies forms stars, it will make
rotationally supported stellar disks with radii of a few kpc.
These stellar disks will probably contain on the order of a percent of
the total stellar mass.
The disks will be very similar in character to the stellar
disks which are now known in many ellipticals \citep{scorza98},
though perhaps somewhat less luminous than the stellar disks which are
detectable at the present time.

\acknowledgments

Thanks to Michael Rupen for helping to get this project off the ground.
Thanks to Tom Statler and the referee, C. Henkel, for helpful
comments.
Thanks also to the Berkeley-Illinois-Maryland Association and the Owens
Valley Radio Observatory (both operated with support from the National
Science Foundation) for generous investments of telescope time.
The Digitized Sky Surveys were produced at the Space Telescope Science
Institute under U.S. Government grant NAG W-2166. 
This publication also makes use of data products from the Two Micron All Sky
Survey, which is a joint project of the University of Massachusetts and
the
Infrared Processing and Analysis Center, funded by the National
Aeronautics and Space Administration and the National Science
Foundation. 
This research has made use of the NASA/IPAC Extragalactic Database (NED) which is operated by the Jet Propulsion Laboratory, California Institute of Technology, under contract with the National Aeronautics and Space Administration. 
Last but not least,
this work has been supported by National Science Foundation grant AST
00-74709.

\end{document}